\documentclass[12pt,a4paper]{amsart}
\usepackage{amssymb,color,pstricks,caption}

\newtheorem{theorem}{Theorem}

\newtheorem{corollary}[theorem]{Corollary}

\newtheorem{lemma}[theorem]{Lemma}

\newcommand{\Proof}{\medskip\noindent\emph{Proof. }}

\newcommand{\cqfd}{\hfill $\Box$\medskip}
\newcommand{\ov}{\overline}
\newcommand{\boldzero}{\boldsymbol{0}}
\newcommand{\boldLambda}{\boldsymbol{\Lambda}}
\newcommand{\boldlambda}{\boldsymbol{\lambda}}
\newcommand{\boldalpha}{\boldsymbol{\alpha}}
\newcommand{\boldbeta}{\boldsymbol{\beta}}
\newcommand{\bolddelta}{\boldsymbol{\delta}}

\newcommand{\qbinom}[2]{{\genfrac{[}{]}{0pt}{}{#1}{#2}}_q}

\def\ZZ{{\mathbb{Z}}}

\def\ZZp{{\mathbb{Z}_{\ge0}}}

\def\ll{\left \lgroup}
\def\rr{\right \rgroup}
\def\pp{p^{\prime}}

\def\cM{\mathcal M}
\def\cO{\mathcal O}
\def\cW{\mathcal W}

\newcommand{\CPX}[2]{{{\mathcal{C}}^{#1}_{#2}}}
\newcommand{\CPgfX}[2]{{C^{#1}_{#2}}}
\newcommand{\HLX}[2]{{{\mathcal H}^{#1}_{#2}}}
\newcommand{\HLgfX}[2]{{H^{#1}_{#2}}}

\newcommand{\B}{\mathcal B}

\newcommand{\lieg}{{\mathfrak{g}}}
\newcommand{\slclass}[1]{{\mathfrak{sl}}_{#1}}
\newcommand{\slchap}[1]{\widehat{\mathfrak{sl}}_{#1}}

\newcommand{\cartan}{{\mathfrak h}}
\newcommand{\cartand}{{\mathfrak h}^*}
\newcommand{\rootlatc}[1]{Q_{#1}}   
\newcommand{\wtlatd}[2]{\ov P^+_{#1,#2}}   
\newcommand{\wtinner}[2]{\langle{#1}|{#2}\rangle}

\newcommand{\partset}[1]{{\mathcal P}_{#1}}

\newcommand{\hbress}{h}
\newcommand{\thbress}{\tilde h}
\newcommand{\hh}{h}
\newcommand{\ho}{h^\circ}
\newcommand{\hp}{h^{+}}
\newcommand{\wt}{\text{wt}}
\newcommand{\npt}{\text{np}}

\newcommand{\parop}{\hat\lambda}
\newcommand{\omegaop}{\hat\omega}

\newcommand{\Adset}[2]{{\mathcal A}^{#1}_{#2}} 
\newcommand{\Adb}{\Adset{k}{b}}
\newcommand{\AGF}[2]{{A}^{#1}_{#2}} 
\newcommand{\AdbGF}{\AGF{k}{b}}
\newcommand{\Bdset}[2]{{\mathcal B}^{#1}_{#2}} 
\newcommand{\Bdb}{\Bdset{k}{b}}
\newcommand{\BGF}[2]{{B}^{#1}_{#2}} 
\newcommand{\BdbGF}{\BGF{k}{b}}

\newcommand{\tAdset}[2]{\tilde{\mathcal A}^{#1}_{#2}} 
\newcommand{\tAdb}{\tAdset{k}{b}}
\newcommand{\tAGF}[2]{{\tilde A}^{#1}_{#2}} 
\newcommand{\tAdbGF}{\tAGF{k}{b}}
\newcommand{\tBdset}[2]{\tilde{\mathcal B}^{#1}_{#2}} 
\newcommand{\tBdb}{\tBdset{k}{b}}
\newcommand{\tBGF}[2]{{\tilde B}^{#1}_{#2}} 
\newcommand{\tBdbGF}{\tBGF{k}{b}}

\long\def\ednull#1{}

\begin{document}
\title[]{Cylindric partitions, $\cW_r$ characters and the Andrews-Gordon-Bressoud identities}
\dedicatory{Dedicated to Prof R J Baxter on his 75th birthday}

\author{O Foda}
\address{School of Mathematics and Statistics,
  University of Melbourne,
  Victoria 3010, Australia}
\email{omar.foda@unimelb.edu.au}

\author{T A Welsh}
\address{School of Mathematics and Statistics,
  University of Melbourne,
  Victoria 3010, Australia}
\email{trevor.welsh@unimelb.edu.au}

\date{}

\begin{abstract}
We study the Andrews-Gordon-Bressoud (AGB) generalisations of the Rogers-Ramanu\-jan $q$-series 
identities in the context of cylindric partitions.  
We recall the definition of $r$-cylindric partitions, and provide a simple proof of Borodin's 
product expression for their generating functions, that can be regarded as a limiting case of 
an unpublished proof by Krattenthaler. 
We also recall the relationships between 
the $r$-cylindric partition generating functions,
the principal characters of $\slchap{r}$ algebras, 
the $\cM^{\, r, r+d}_r$ minimal model characters of $\cW_r$ algebras, and 
the $r$-string abaci generating functions, as well as 
the relationships between them, providing simple proofs for each.    

We then set $r\!=\!2$, and use $2$-cylindric partitions to re-derive the AGB identities as follows. 
Firstly, we use Borodin's product expression for the generating functions of the 2-cylindric 
partitions with infinitely-long parts, to obtain the product sides of the AGB identities, 
times a factor $(q; q)_{\infty}^{-1}$, which is the generating function of ordinary partitions.   
Next, we obtain a bijection from the 2-cylindric partitions, {\it via} 2-string abaci, into 
decorated versions of Bressoud's restricted lattice paths. Extending Bressoud's method of 
transforming between restricted paths that obey different restrictions, we obtain sum expressions 
with manifestly non-negative coefficients for the generating functions of the 2-cylindric partitions 
which contains a factor $(q; q)_{\infty}^{-1}$. Equating the product and sum expressions of the same 
2-cylindric partitions, and canceling a factor of $(q; q)_{\infty}^{-1}$ on each side, we obtain 
the AGB identities.  
\end{abstract}

\maketitle

\section{Introduction}
\label{introduction}

{\it 
We recall how the Rogers-Ramanu\-jan identities and their Andrews-Gordon extensions appear in statistical 
mechanical models in terms of counting weighted paths. 
Next, we motivate our re-derivation of these identities in terms of counting cylindric partitions. 
}

\subsection{Baxter's corner transfer matrices and Rogers-Ramanu\-jan identities} One of R J Baxter's 
many profound contributions to statistical mechanics is the corner transfer matrix method 
\cite{baxter.ctm.01}--\cite{baxter.ctm.05}, 
which he introduced to compute order parameters in two-dimensional statistical mechanical lattice 
models with local weights that satisfy Yang-Baxter equations \cite{baxter.book}. 
One of the surprising consequences of the corner transfer matrix method is the appearance of weighted
paths whose generating functions coincide with the very expressions that appear on either side of the
Rogers-Ramanu\-jan $q$-series identities, and generalisations thereof, in the context of computing 
order parameters \cite{baxter.rr}. 

\subsection{Restricted solid-on-solid models} 
An important class of statistical mechanical models are the restricted solid-on-solid models based 
on the Virasoro algebra $\cW_2$ \cite{andrews.baxter.forrester, forrester.baxter}, and their 
higher-rank extensions based on $\cW_r$, $r=3, 4, \cdots$ \cite{jimbo.miwa.okado.01, jimbo.miwa.okado.02}
\footnote{\,
There are {\it \lq higher-level\rq\,} versions of the restricted solid-on-solid models based on $\cW_2$,
obtained by a fusion procedure \cite{date.jimbo.kuniba.miwa.okado.01, date.jimbo.kuniba.miwa.okado.02}. 
The same applies, at least in principle, to all restricted solid-on-solid models based on higher rank 
$\cW_r$. 
The simplest examples of the fused $\cW_2$ models are related to the $N\!=\!1$ supersymmetric Virasoro 
algebra. These models, or at least the conformal field theories obtained in the critical limits thereof,  
are relevant to the Bressoud identities discussed in section {\bf \ref{bressoud}} 
\cite{berkovich.mccoy.orrick.1996, berkovich.mccoy.1997}. In this work, we approach the Bressoud identities 
in a combinatorial way.
}.
These models are labelled by two coprime integers $p$ and $\pp$, such that $1 < p < \pp$. We denote the models 
based on $\cW_r$, and labelled by $p$ and $\pp$, by $\cM_r^{\, p, \pp}$. In this work, we focus on the models 
$\cM^{\, r, \pp}_r$, that is $p=r$, then restrict further to $\cM^{\, 2, 2k+3}_2$, $k=1, 2, \cdots$

\subsection{Order parameters and Rogers-Ramanujan-type identities} 
An important set of observables in statistical mechanical models is the set of {\it \lq order parameters\rq\,} 
\footnote{
The order parameters are also known as {\it \lq local height probabilities\rq\,}.
}. 
In the following, we denote the order parameters of the models $\cM^{\, 2, 2k+3}_2$, $k=1, 2, \cdots$, by 
$\cO^{\, 2, 2k+3}_s$, $s = 0, 1, \cdots, k$ \cite{forrester.baxter}
\footnote{\,
In all generality, $\cW_2$ restricted solid-on-solid models are labelled by all possible pairs of 
coprime integers $p$ and $\pp$, such that $p < \pp$. Their order parameters, or more generally, 
the primary fields in the corresponding conformal field theories, are labelled by two integers $r_1$ 
and $s_1$, typically chosen such that $r_1 = 1, 2, \cdots, p-1$, and $s_1 = 1, 2, \cdots, \pp -1$. 
When $p=2$, $r_1$ is fixed, $r_1 = 1$, and only $s_1$ varies. The $\cW_2$ restricted solid-on-solid models 
have a $\ZZ_2$-symmetry that reduces the number of independent order parameters. In particular, 
in the models $\cM^{\, 2, 2k+3}_2$, $k=1, 2, \cdots$, $s_1$ is further restricted to $s_1 = 1, 2, \cdots, k+1$.
In this work, we choose to parameterise the order parameters using $s = s_1 - 1 = 0, 1, \cdots, k$,
which is better suited for our purposes, as it represents the height of the starting point of 
the 1-dimensional paths that are used to compute $\cO^{\, 2, 2k+3}_s$. 
}. 
The Andrews-Gordon identities follow from deriving two expressions for $\cO^{\, 2, 2k+3}_s$, for each 
$k \ge 1$ and $0 \le s \le k$, one in a sum form with manifestly non-negative coefficients
\footnote{\,
In the sequel, we refer to these sum forms as {\it \lq non-negative sum forms\lq\,}.
}, and one in product form, and equating them.

\subsection{Two-dimensional lattice configurations}
To compute $\cO^{\, 2, 2k+3}_s$, one needs to sum over all possible two-dimensional lattice configurations, 
such that the state variable in the middle of the lattice is fixed to a specific value that we parameterise 
in terms of $s$, and the state variables on the boundaries of the lattice, sufficiently far away from the 
middle of the lattice, are in ground-state positions. There is only one ground state in
each $\cM^{\, 2, 2k+3}_2$ model \cite{forrester.baxter}.
This is a difficult problem not only because it requires summing over two-dimensional lattice configurations,
but also because these configurations are weighted by products of local weights that are typically complicated 
functions of the rapidity variables. 

\subsection{One-dimensional paths on restricted one-dimensional lattices} 
\label{one.dimensional.paths.on.restricted.one.dimensional.lattices}
In the corner transfer matrix approach to the order parameters in $\cM^{\, p, \pp}_2$ models, the Yang-Baxter 
equations are used to reduce the sums over two-dimensional lattice configurations with complicated local weights, 
to sums over one-dimensional paths, on restricted one-dimensional lattices, that is one-dimensional lattices with 
finitely-many sites. As we will see in later sections, these one-dimensional paths have simple local weights. 

In the case of $\cM^{\, 2, 2k+3}_2$, the restricted one-dimensional lattice has $k+1$ sites labelled by 
$s=0, 1, \cdots, k$.  The set of paths that are used to compute $\cO^{\, 2, 2k+3}_s$, where 
$k=1, 2, \cdots$, and $s=0, 1, \cdots, k$ is denoted $\Adset{k}{s}$. 

\subsection{The paths as walks on a discrete lattice in discrete time}
One can think of the paths that are elements of $\Adset{k}{s}$ as
walks on a one-dimensional $(k+1)$-site lattice, in discrete time $t$, such that  
{\bf 1.} each path starts, at $t=0$, at site $s$, 
{\bf 2.} if a path is at site $h \in \{1, 2, \cdots, k-1\}$ at time $t$, then it must be at site 
$h^{\prime} = h \pm 1$ at time $t+1$, 
{\bf 3.} if a path is at site $h=k$, at time $t$, then it must be at $h^{\prime} = k-1$, at time $t+1$, 
{\bf 4.} if a path is at site $h=0$, at time $t$, then it can be at $h=0$ or at $h=1$, at time $t+1$, 
that is, a path is allowed to stay at $h=0$ for as long as it wishes, and finally
{\bf 5.} a path must asymptotically settle at $h=0$ and stay at $h=0$
at all $t > N$, for large but finite $N$.

\subsubsection{Example} The model $\cM^{\, 2, 5}_2$ has two order parameters $\cO^{\, 2, 5}_0$ and 
$\cO^{\, 2, 5}_1$. An example of a path in the set $\Adset{1}{0}$, whose weighted generating function 
leads to the order parameter $\cO^{\, 2, 5}_0$, is in Figure {\bf \ref{A.simple.path}}.

\begin{figure}[ht]
\begin{center}
\psset{yunit=0.64cm,xunit=0.54cm}
\begin{pspicture}(-2,-1)(32,2.5)
\rput[bl](0,0){
\psset{linewidth=0.25pt,linestyle=dashed, dash=2.5pt 1.5pt,linecolor=gray}
\multips(0,0)(0,0){ 1}{\psline{-}(0,0)(24,0.0)}
\multips(0,1)(0,1){ 1}{\psline{-}(0,0)(24,0.0)}
\multips(0,0)(0,0){24}{\psline{-}(0,0)( 0,1.0)}
\multips(1,0)(1,0){24}{\psline{-}(0,0)( 0,1.0)}
\psset{linewidth=0.25pt,fillstyle=none,linestyle=solid,linecolor=black}
\rput[r](-0.5,1){\scriptsize $1$}
\rput[r](-0.5,0){\scriptsize $0$}
\psset{linewidth=0.7pt,fillstyle=none,linestyle=solid,linecolor=black}
\psline( 0,0)( 1,1)( 2,0)( 3,1)( 4,0)( 5,0)( 6,0)( 7,1)( 8,0)( 9,0)
       (10,1)(11,0)(12,1)(13,0)(14,0)(15,0)(16,0)(17,1)(18,0)(19,1)
       (20,0)(21,0)(22,0)(23,0)(24,0)
}
\end{pspicture}
\end{center}
\caption{\it 
A path in $\Adset{1}{0}$.
The vertical axis is the 2-site one-dimensional lattice that the paths are walks on.  
The horizontal axis represents the discrete time $t$ that starts at $t=0$. The paths 
starts at $h=0$ because it is in $\Adset{1}{0}$.
}
\label{A.simple.path}
\end{figure}

\subsection{Alternating-sign $q$-series expressions} The generating functions of 
the one-dimensional paths with simple weights, obtained using the corner transfer matrix 
method, can be evaluated relatively straightforwardly using inclusion-exclusion, in the 
form of alternating-sign $q$-series \cite{andrews.baxter.forrester, forrester.baxter}. 

\subsubsection{Pochhammer symbol}
We need the following standard definition of the Pochhammer symbol $(z;q)_n$ and its limit 
$(z;q)_\infty$,

\begin{equation}
(z; q)_0 = 1, 
\quad
(z; q)_n = \prod_{i=0}^{n-1}(1-zq^i), 
\quad
(z;q)_\infty=\prod_{i=0}^{\infty}(1-zq^i)
\end{equation}

\subsubsection{Example} In $\cM^{\, 2, 5}_2$, there are two allowed values $s$ for the height 
of the initial point of the paths, namely $s=0, 1$, 
and one obtains 

\begin{equation}
\label{RR.alternating}
\cO^{2, 5}_s (q) \sim \frac{1}{ (q; q)_{\infty} } 
\sum_{n=-\infty}^{\infty} 
\ll
q^{    n   (10 n + 1 + 2s) } - 
q^{  (2n+1)( 5 n + 2 -  s) } 
\rr,
\quad s = 0, 1,
\end{equation}

\noindent where we have used the proportionality sign $\sim$ to indicate that we have neglected 
a normalisation constant that need not concern us here. 

\subsubsection{Bosonic expressions}
In all generality, the order parameters in $\cM^{\, p, \pp}_r$ can be evaluated in alternating-sign
series form. These alternating-sign series coincide with the characters of the minimal $\cW_r$ 
conformal field theories labelled by coprime $p$ and $\pp$, when these characters are evaluated 
in Feigin-Fuchs form \cite{feigin.fuchs.1982, rocha.caridi.1984}. These forms are called {\it bosonic} 
\cite{bosonic.fermionic.01}--\cite{bosonic.fermionic.03}, since they 
can be obtained using the oscillators of $(r-1)$ free boson fields, in a background charge.

\subsection{Manifestly non-negative $q$-series expressions} Since the generating functions, computed 
using the corner transfer matrix method, count weighted paths, it is natural to try to re-write them 
as $q$-series with non-negative coefficients. This turns out to be less straightforward 
to do, compared to the derivation of the alternating-sign expression, but can be done in many cases. 
In the following, we refer to the non-negative-sign expression as {\it sum expressions}. 

\subsubsection{Fermionic expressions}
The non-negative sum expressions provide other forms for the corresponding $\cW_2$ characters.
These forms are called {\it fermionic}
\cite{bosonic.fermionic.01}--\cite{bosonic.fermionic.07}.

\subsubsection{Example} The alternating-sign expressions on the right hand side of \eqref{RR.alternating} 
can be put in non-negative sum form. One way to do this is to compute the generating function 
of the weighted paths combinatorially \cite{bressoud.1989}, to obtain

\begin{equation}
\label{RR.sum}
\frac{1}{ (q; q)_{\infty} }
\sum_{n=-\infty}^{\infty}
\ll
q^{    n   (10 n + 1 + 2s) } -
q^{  (2n+1)( 5 n + 2 -  s) }
\rr
= \sum_{n=0}^{\infty} \frac{ q^{n^2 + sn} }{ (q; q)_n }, \quad s = 0, 1,
\end{equation}

\noindent But these sum expressions are nothing but the sum sides of
the celebrated Rogers-Ramanu\-jan identities.

\subsection{Product $q$-series expressions and Rogers-Ramanujan-type identities}
In the case of the order parameters $\cO^{\, 2, 2k+3}_2$, $k=1, 2, \cdots$, one 
can use $q$-series identities to put the sum expressions in product form 
\footnote{\,
To be precise, one uses the $q$-series identity to put the alternating-sign sum 
side in product form, then uses the equality of the alternating-sign form and 
the non-negative form, proven separately.}.
Equating the non-negative sum and the product expressions, one obtains the 
Rogers-Ramanu\-jan identities and the Andrews-Gordon extensions thereof.

\subsubsection{Example} The sum expressions on the right hand side of \eqref{RR.sum}, 
can be put in product form as

\begin{equation}
\label{RR.product}
\sum_{n=0}^{\infty} \frac{ q^{n^2 + sn} }{ (q; q)_n }
= 
\prod_{\begin{subarray}{c} n=1 \\ n \equiv \pm (s+1) \, \textit{mod} \, 5 \end{subarray}}^{\infty}
\frac{1}{1-q^n}, 
\quad s = 0, 1,
\end{equation}

\noindent which are the original Rogers-Ramanu\-jan identities. 

\subsection{Purpose of this work}
The Rogers-Ramanujan identities and their Andrews-Gordon extensions are related to the Virasoro algebra 
$\cW_2$.
The Bressoud identities are related to the $N\!=\! 1$ supersymmetric extension of the Virasoro algebra. 
They can be regarded as equalities between sum and product expressions for the generating 
functions of one-dimensional weighted paths. 
For models based on $\cW_r$ algebras, $r = 3, 4, \cdots$, paths are available, but 
they are walks on restricted $(r-1)$-dimensional lattices, and therefore are not as useful, and as 
easy to work with as in the $\cW_2$ case. 

We re-derive the AGB identities as equalities between sum and product expressions for the generating 
functions of 2-cylindric partitions. 
Since the latter have useful higher-rank analogues, this approach may lead to a systematic derivation 
of the higher rank identities
\footnote{\,
As we will see in section {\bf \ref{cylindric.partitions}}, the $\cW_r$ product sides are known from 
the $r$-cylindric partition approach, for all $r$.
The $\cW_3$ sum sides are known from \cite{andrews.schilling.warnaar.1999, feigin.foda.welsh.2008}. 
The $\cW_r$ sum sides, for $r=4, 5, \cdots$, are yet to be found. The purpose of this work is to 
pave a possible way to compute them.
}.

\subsection{Outline of contents}
In section {\bf \ref{cylindric.partitions}}, we recall basic facts related to $r$-cylindric partitions,
then provide a straightforward proof of Borodin's product expression for their generating functions.
In section {\bf \ref{characters}}, we recall basic facts related to the principal characters of $\slchap{r}$ 
algebras, and the $\cM^{\, r, r+d}_r$ minimal model characters of $\cW_r$.
In section {\bf \ref{abacus}}, we express the $r$-cylindric partitions in terms of $r$-string abaci.
In section {\bf \ref{andrews.gordon.bressoud}}, we state our results for both the Andrews-Gordon and
the Bressoud identities.
In section {\bf \ref{paths.decorated.paths.and.transforms}}, we introduce Bressoud's paths, the related 
decorated paths, and Bressoud's transforms between different paths.
In section {\bf \ref{andrews.gordon}}, we focus on the $r=2$ case, use the 2-string abaci to map the 
2-cylindric partitions to decorated Bressoud-type paths, then use the Bressoud transforms between paths 
to obtain sum expressions for the generating functions of the $r$-cylindric partitions. Equating the
product expressions of section {\bf \ref{cylindric.partitions}} with the sum expressions of this 
section, we obtain the Andrews-Gordon identities.
Section {\bf \ref{translation}} reconsiders the proof in section {\bf \ref{andrews.gordon}} in terms
of abaci.
In {\bf \ref{bressoud}}, we prove the Bressoud identities.
Section {\bf \ref{discussion}} includes comments and remarks.
Three appendixes contain reviews of technical details. 
Appendix {\bf \ref{macdonald.identity}} recalls the $\slchap{r}$ Macdonald identity.
Appendix {\bf \ref{affine.characters}} contains basic facts related to $\slchap{r}$ weight space 
and characters.
Appendix {\bf \ref{Kyoto}} is on the Kyoto patterns and their relation to cylindric partitions. 

\section{Cylindric partitions}
\label{cylindric.partitions}

\noindent {\it 
We recall basic facts related to cylindric partitions, and provide a simple proof of Borodin's
product expression for their generating functions.
}

\subsection{Brief history} 
\label{brief.history} 
Two-row, that is, 2-cylindric partitions were introduced, using different terminology,
by Burge \cite{burge.1993}. They were formally defined in general form and studied by
Gessel and Krattenthaler \cite{gessel.krattenthaler.1997}.
Special cases were already present in the work of the Kyoto group
\cite{date.jimbo.kuniba.miwa.okado.1989,jimbo.misra.miwa.okado.1991}
on the combinatorics of representations of $\slchap{r}$ and
its quantum analogue $U_q(\slchap{r})$ at $q=0$.
In another guise, they were used to label the irreducible representations
of the Ariki-Koike algebras for 
certain choices of the parameters of these algebras \cite{ariki.1996}.

\subsection{Cylindric partitions as restricted plane partitions} What we refer to, in section 
{\bf \ref{introduction}}, as $r$-cylindric partitions, are plane partitions that consist of $r$ rows, 
with a condition that relates the entries in the $1$-st and the $r$-th rows. Because of this condition, 
one can naturally draw cylindric partitions on a cylinder. 

\subsection{Cylindric partitions with infinitely-long rows} 
\label{cylindric.partitions.with.infinitely.long.rows} 
The special cases mentioned in subsection 
{\bf \ref{brief.history}} are such that the rows are allowed to be infinitely long. In this work, we will 
focus on the cylindric partitions with possibly infinitely-long rows.

\subsection{Borodin's product expression and relation to affine and $\cW_r$ algebras} A product expression 
for the generating functions of the $r$-cylindric partitions with possibly infinitely-long rows, was obtained 
by Borodin \cite{borodin.2007}. This product expression shows that the generating functions of these 
$r$-cylindric partitions are closely related to the principally-specialised characters of integrable 
representations of the affine Lie algebra $\slchap{r}$, or equivalently $A^{(1)}_{r-1}$, as well as to 
characters of specific minimal conformal field theories based on, or equivalently, specific degenerate 
representations of the $\cW_r$ algebras, the $r=2$ case of which is the Virasoro algebra. If $\Lambda$ 
is a level $d$ dominant integral weight of $\slchap{r}$, then there is a set of cylindric partitions 
associated with $\Lambda$, as will be made clear below, whose generating function $\CPgfX{r}{\Lambda}$ 
satisfies

\begin{equation}
\label{Eq:Chars}
\CPgfX{r}{\Lambda}=
\frac{\text{Pr}\,\chi^{\slchap{r}}_\Lambda}
                    {(q^r;q^r)_\infty}
=\frac{\chi^{\cW_r}_\Lambda}{(q;q)_\infty},
\end{equation}

\noindent where $\text{Pr}\,\chi^{\slchap{r}}_\Lambda$ is the principal character of $\slchap{r}$ labelled 
by $\Lambda$, and $\chi^{\cW_r}_\Lambda$ is a normalised character of the $\cW_r$ minimal model 
$\cM^{\, r, r+d}_r$, this character also labelled by $\Lambda$.

\noindent Note that the minimal models $\cM^{\, r, r+d}_r$,
based on the algebras $\cW_r$, are defined only 
if $r$ and $d$ are coprime
\footnote{\,
As mentioned above, we will not discuss statistical mechanical models, or conformal field theories, based 
on supersymmetric extensions of $\cW_r$, where the labels $p$ and $\pp$ need not be co-prime, in this work.
Instead, we will obtain the corresponding characters, combinatorially, from those related to $\cW_r$.
}.
Thus in other than such a case, the second equality in \eqref{Eq:Chars} should be ignored. However, the first 
equality continues to hold.

\subsection{Ordinary partitions}
\label{ordinary.partitions}
We define an ordinary partition $\lambda$ to be a sequence $\lambda=(\lambda_1, \lambda_2$, $\cdots)$ for which 
$\lambda_1\ge\lambda_2\ge \cdots\ge0$, with only finitely-many non-zero elements. If we write 
$\lambda = (\lambda_1, \cdots, \lambda_p)$, then we imply that $\lambda_i = 0$ for $i > p$. The length 
$\ell(\lambda)$ of $\lambda$ is the smallest $\ell$ such that $\lambda_{\ell+1}=0$, while the weight 
$|\lambda|$ of $\lambda$ is defined by $|\lambda| = \lambda_1 + \lambda_2 + \cdots + \lambda_{\ell(\lambda)}$.

\subsection{Formal definition of cylindric partitions}
\label{Cyl}

Following Gessel and Krattenthaler \cite{gessel.krattenthaler.1997}, given two ordinary partitions 
$\lambda=(\lambda_1,\lambda_2, \cdots, \lambda_r)$ and $\mu=(\mu_1, \mu_2, \cdots, \mu_r)$, and $d \ge 0$,
a cylindric partition of type $\lambda / \mu / d$ is an array of non-negative integers $\pi_{ij}$ of the form

\begin{equation}\label{Eq:PlaneArray}
\begin{array}{cccccccccccc}
&&&&&\pi_{1,\mu_1+1}& \cdots& \cdots& \cdots& \cdots& \cdots&\pi_{1,\lambda_1}\\
&&&\pi_{2,\mu_2+1}& \cdots&\pi_{2,\mu_1+1}& \cdots& \cdots& \cdots& \cdots
&\pi_{2,\lambda_2}\\
&&&\vdots&&\vdots&&&\ddots\\
&\pi_{r,\mu_r+1}& \cdots& \cdots& \cdots& \cdots& \cdots&\pi_{r,\lambda_r}\\
\end{array}
\end{equation}

\noindent such that the entries are weakly decreasing across the rows and
down the columns, {\it i.e.},

\begin{subequations}\label{Eq:PlaneCons}
\begin{align}\label{Eq:PlaneRow1}
\pi_{i,j}&\ge\pi_{i,j+1}
&
(1\le i\le r,\,\mu_i\le j<\lambda_i),\\
\label{Eq:PlaneRow2}
\pi_{i,j}&\ge\pi_{i+1,j}
&
(1\le i<r\,,\max\{\mu_i,\mu_{i+1}\}\le j\le\min\{\lambda_i,\lambda_{i+1}\}),\\
\noalign{\noindent and also respect the cylindrical condition}
\label{Eq:PlaneCyl}
\pi_{r,j}&\ge\pi_{1,j+d}
&
(\max\{\mu_r,\mu_1-d\}\le j\le\min\{\lambda_r,\lambda_1-d\})
\end{align}
\end{subequations}

\noindent Condition \eqref{Eq:PlaneCyl} can be regarded as the weakly decreasing column condition,
extending to a copy of the first row of the array \eqref{Eq:PlaneArray}, displaced $d$ positions to 
the left and placed below the $r$-th row.
Note that the ordinary partition $\lambda$  has $r$ parts, 
and these parts determine the maximal possible lengths of the $r$ rows of the $r$-cylindric partition
of type $\lambda / \mu / d$. 
On the other hand, 
the ordinary partition $\mu$ also has $r$ parts, 
and these parts determine the relative shifts          of the $r$ rows of the $r$-cylindric partition
of type $\lambda / \mu / d$. 

As mentioned in subsection {\bf \ref{cylindric.partitions.with.infinitely.long.rows}}, in this paper, 
we are primarily interested in cylindric partitions with rows that are allowed to be arbitrarily long. 
These cylindric partitions are then of type $(\infty^r)/\mu/d$.

\subsubsection{Example} 
\label{example.01}
A cylindric partition of type $(\infty^r)/\mu/d$, such that $r=3$, $\mu=(1,1,0)$,
and $d=4$, is

\begin{equation}\label{Eq:TypCPP1}
\begin{tabular}{ccccccccccccc}
&5&4&4&2&2&1&1&$\cdot$&$\cdot$&$\cdot$\\
&3&2&2&2&1&$\cdot$&$\cdot$&$\cdot$&$\cdot$&$\cdot$\\
4&2&2&1&$\cdot$&$\cdot$&$\cdot$&$\cdot$&$\cdot$&$\cdot$&$\cdot$
\end{tabular}
\end{equation}
where each dot represents a zero, and these are assumed to extend
indefinitely to the right.

\subsection{Working on a cylinder}
As mentioned above, it is useful to visualise the cylindric condition \eqref{Eq:PlaneCyl} by placing a copy 
of the first row, displaced $d$ positions to the left, below the $r$-th row. In fact, it is useful to place 
copies of the whole cylindric partition above and below the {\it \lq main copy\rq\,}, so to speak, displaced 
appropriately. 

\subsubsection{Example}
\label{example.02}
From the example in paragraph {\bf \ref{example.01}}, we obtain,

\begin{equation}\label{Eq:TypCPP2}
{\scriptsize
\begin{tabular}{ccccccccccccccccccc}
&&&&&&&&&5&4&4&2&2&1&1&$\cdot$&$\cdot$&$\cdot$\\
&&&&&&&&&3&2&2&2&1&$\cdot$&$\cdot$&$\cdot$\\
&&&&&&&&4&2&2&1&$\cdot$&$\cdot$&$\cdot$\\
&&&&&5&4&4&2&2&1&1&$\cdot$&$\cdot$&$\cdot$\\
&&&&&3&2&2&2&1&$\cdot$&$\cdot$&$\cdot$\\
&&&&4&2&2&1&$\cdot$&$\cdot$&$\cdot$\\
&5&4&4&2&2&1&1&$\cdot$&$\cdot$&$\cdot$\\
&3&2&2&2&1&$\cdot$&$\cdot$&$\cdot$\\
4&2&2&1&$\cdot$&$\cdot$&$\cdot$
\end{tabular}
}
\end{equation}

\subsection{Level-rank duality} 
\label{level.rank.duality}
By focussing on $d$ consecutive columns of this array, we see that cylindric 
partitions are in bijection with those having $r$ and $d$ interchanged, and $\mu$ replaced by its conjugate 
$\mu^{\prime}$. This demonstrates a \emph{level-rank duality}.

\subsection{Labeling cylindric partitions by affine fundamental weights}
Let $\Lambda_0, \Lambda_1, $ $\cdots,$ $\Lambda_{r-1}$ denote the fundamental weights of $\slchap{r}$. The 
level-$d$ dominant integral weight lattice of $\slchap{r}$ is then defined by 

\begin{equation}
\wtlatd{r}{d} = \sum_{i=0}^{r-1}m_i\Lambda_i| m_i \in \ZZp, \quad \text{for} \quad 0 \le i < r, 
                                                            \quad \sum_{i=0}^{r-1} m_i = d
\end{equation}

\noindent A weight $\sum_{i=0}^{r-1}m_i\Lambda_i$ will be denoted $[m_0, m_1, \cdots, m_{r-1}]$.

Let $\partset{k}$ denote the set of all partitions $\lambda$ for which $\ell(\lambda)\le k$.
For $\Lambda=\sum_{i=0}^{r-1}m_i\Lambda_i\in\wtlatd{r}{d}$, we define a corresponding partition
$\parop(\Lambda)=(\mu_1,\mu_2, \cdots)$ by setting $\mu_j=\sum_{i=j}^{r-1}m_i$ for $j=1, \cdots,r-1$
and $\mu_j=0$ for $j\ge r$.  Thus $\parop(\Lambda)\in\partset{r-1}$.

\subsection{Generating functions}
For $\Lambda\in\wtlatd{r}{d}$ and $\mu=\parop(\Lambda)$,
define $\CPX{r}{\Lambda}$ to be the set of all
cylindric partitions of type $(\infty^r)/\mu/d$
that have a finite number of non-zero entries,
and let $\CPX{r}{\Lambda}(a)$ be the subset of $\CPX{r}{\Lambda}$
comprising those $\pi$ whose entries don't exceed $a$.
For $\pi\in\CPX{r}{\Lambda}$, define the norm $|\pi|$ to be the
sum of the entries in $\pi$.
In the case of the cylindric partition $\pi$ of
\eqref{Eq:TypCPP1} and \eqref{Eq:TypCPP2}, $|\pi|=38$.
Define the generating functions

\begin{equation}\label{Eq:CPgfDef}
\CPgfX{r}{\Lambda}=\sum_{\pi\in\CPX{r}{\Lambda}} q^{|\pi|},\qquad
\CPgfX{r}{\Lambda}(a)=\sum_{\pi\in\CPX{r}{\Lambda}(a)} q^{|\pi|}
\end{equation}

\subsection{Product formula}

There is a remarkably simple product formula for $\CPgfX{r}{\Lambda}$, first derived by Borodin \cite{borodin.2007}.
To express this, first construct a length $(r+d)$ binary word
$\omegaop(\Lambda) = \omega_1 \omega_2 \cdots \omega_{r+d}$ from $\Lambda\in\wtlatd{r}{d}$.
This word comprises $r$ $0$'s and $d$ $1$'s, and for $\Lambda=\sum_{i=0}^{r-1}m_i\Lambda_i$, is defined by putting 
the $j$-th $0$ at position $j+\sum_{i=0}^{j-1}m_i$.

Alternatively, consider the partition $\mu$ as sitting in an $r\times d$ box. Then on viewing the profile of $\mu$ 
as a lattice path passing from the top right of the box to the bottom left, the word $\omegaop(\Lambda)$ is 
constructed by interpreting each horizontal step as $1$ and each vertical step as $0$.

\begin{theorem}\label{CPPprodbin}
\cite[Proposition 5.1]{borodin.2007}
If $\Lambda\in\wtlatd{r}{d}$ and $\omega=\omegaop(\Lambda)$,
then
\begin{equation}\label{Eq:CPPprodbin}
\CPgfX{r}{\Lambda}
=\frac{1}{(q^{r+d};q^{r+d})_\infty}
\prod_{\begin{subarray}{c}
        1\le i<j\le r+d\\ \omega_i>\omega_j
      \end{subarray}}
\frac{1}{(q^{j-i};q^{r+d})_\infty}
\prod_{\begin{subarray}{c}
        1\le i<j\le r+d\\ \omega_i<\omega_j
      \end{subarray}}
\frac{1}{(q^{r+d-j+i};q^{r+d})_\infty}
\end{equation}
\end{theorem}

\noindent Below, we provide a straightforward proof of this result, based on the results of 
\cite{gessel.krattenthaler.1997}. This proof may be viewed as a limiting case of an unpublished 
proof of Krattenthaler \cite{krattenthaler.2008}.
Before doing so, we illustrate the result, demonstrating that it may be cast in terms of 
\emph{hook-lengths}. 

\subsection{Hook lengths}
The hook-length of a node is the number of nodes in the hook comprising the nodes directly above 
and to the left of the given node.

\subsection{The product expression in terms of hook lengths}
Consider the case $r=5$ with $\Lambda=[1,3,0,2,1]$ so that $d=7$.
From this we obtain $\omega=\omegaop(\Lambda)=101110011010$. The profile of the corresponding 
$\mu=(6, 3, 3, 1)$ is then the thick line in Figure {\bf \ref{TypicalHookProd}}. The exponents 
$(j-i)$ that appear in the first product on the right side of \eqref{Eq:CPPprodbin} are inserted 
in the boxes below the profile in Figure {\bf \ref{TypicalHookProd}}, and are seen to correspond 
to hook-lengths. The exponents $(r+d-j+i)$ that appear in the second product are inserted in the 
boxes above the profile, and are seen to correspond to {\it \lq reverse\rq\,} hook-lengths.

\begin{figure}[ht]
\begin{center}
\psset{yunit=0.4cm,xunit=0.4cm}
\begin{pspicture}(-2,-7)(10,0)
\rput[bl](0,-7){
\psset{linewidth=0.25pt,fillstyle=none,linestyle=solid,linecolor=black}
\multips(0,0)(0,1){6}{\psline{-}(0,0)(7,0)}
\multips(0,0)(1,0){8}{\psline{-}(0,0)(0,5)}
\psset{linewidth=1.5pt,fillstyle=none,linestyle=solid,linecolor=black}
\psline(0,0)(0,1)(1,1)(1,2)(3,2)(3,4)(6,4)(6,5)(7,5)
\scriptsize
\rput[c](0.5,0.5){$1$}
\rput[c](1.5,0.5){$3$}
\rput[c](2.5,0.5){$4$}
\rput[c](3.5,0.5){$7$}
\rput[c](4.5,0.5){$8$}
\rput[c](5.5,0.5){$9$}
\rput[c](6.5,0.5){$11$}
\rput[c](0.5,1.5){$11$}
\rput[c](1.5,1.5){$1$}
\rput[c](2.5,1.5){$2$}
\rput[c](3.5,1.5){$5$}
\rput[c](4.5,1.5){$6$}
\rput[c](5.5,1.5){$7$}
\rput[c](6.5,1.5){$9$}
\rput[c](0.5,2.5){$8$}
\rput[c](1.5,2.5){$10$}
\rput[c](2.5,2.5){$11$}
\rput[c](3.5,2.5){$2$}
\rput[c](4.5,2.5){$3$}
\rput[c](5.5,2.5){$4$}
\rput[c](6.5,2.5){$6$}
\rput[c](0.5,3.5){$7$}
\rput[c](1.5,3.5){$9$}
\rput[c](2.5,3.5){$10$}
\rput[c](3.5,3.5){$1$}
\rput[c](4.5,3.5){$2$}
\rput[c](5.5,3.5){$3$}
\rput[c](6.5,3.5){$5$}
\rput[c](0.5,4.5){$3$}
\rput[c](1.5,4.5){$5$}
\rput[c](2.5,4.5){$6$}
\rput[c](3.5,4.5){$9$}
\rput[c](4.5,4.5){$10$}
\rput[c](5.5,4.5){$11$}
\rput[c](6.5,4.5){$1$}
}
\end{pspicture}
\end{center}
\caption{\it 
Hook-lengths for $r=5$, $\Lambda=[1,3,0,2,1]$
}
\label{TypicalHookProd}
\end{figure}

\noindent Therefore, we obtain

\begin{equation}\label{Eq:CPPprodbinEx}
\CPgfX{5}{[1,3,0,2,1]}
=\frac{1}{
(q^{12};q^{12})_\infty
(q,q^3,q^9,q^{11};q^{12})^4_\infty
(q^2,q^5,q^6,q^7,q^{10};q^{12})^3_\infty
(q^4,q^8;q^{12})^2_\infty
}\,,
\end{equation}

\noindent where
$(z_1, \cdots,z_\ell;q)_\infty=(z_1;q)_\infty \cdots(z_\ell;q)_\infty$.
Note that it is also possible to place all $(r\times d)$ exponents
below the profile, with all corresponding to genuine hook-lengths,
although they will no longer form a rectangle, in general.

\subsection{Short proof of Theorem {\bf \ref{CPPprodbin}}}

In \cite{andrews.schilling.warnaar.1999}, Andrews, Schilling and Warnaar observed that all $\cM^{\, r, r+d}_r$ 
characters are expressible in product form. This suggests that Theorem {\bf \ref{CPPprodbin}} can be proved 
by applying the $\slchap{r}$ Macdonald identity to the appropriate limit of Theorem 2 of 
\cite{gessel.krattenthaler.1997}.
To apply this latter result to the cylindric partitions of type $(\infty^r)/\mu/d$ defined above, we 
use the case of Theorem 2 of \cite{gessel.krattenthaler.1997} for which each $a_j=a$ and each $b_j=0$,
and then take the limit $\lambda\to(\infty^r)$.

\begin{theorem} \cite[specialisation of Theorem 2]{gessel.krattenthaler.1997}
If $\Lambda\in\wtlatd{r}{d}$ and $\mu=\parop(\Lambda)$, then

\begin{equation}\label{Eq:CPPgenr}
\CPgfX{r}{\Lambda}(a)=
\sum_{k_1+ \cdots+k_r=0}
\det_{1\le s,t\le r}
\ll 
\frac{
q^{(\mu_t-t)(rk_s+s-t)+k_s(\frac12 r k_s + s)(r+d)}
}{
(q;q)_{a+rk_s+s-t}
}
\rr
\end{equation}
\end{theorem}

\noindent By taking the $a\to\infty$ limit in \eqref{Eq:CPPgenr}, we obtain

\begin{equation}\label{Eq:CPPgenrlim}
\CPgfX{r}{\Lambda}
=\lim_{a\to\infty} \CPgfX{r}{\Lambda}(a)
=\frac{1}{(q;q)^r_\infty}
\sum_{k_1+ \cdots+k_r=0}
\det_{1\le s,t\le r}
\ll q^{(\mu_t-t)(rk_s+s-t)+k_s(\frac12 r k_s + s)(r+d)} \rr
\end{equation}

\noindent The $\slchap{r}$ Macdonald identity can be expressed in the form
\footnote{\, 
See Appendix {\bf \ref{macdonald.identity}}
},

\begin{equation}\label{Eq:MacId}
\sum_{k_1+ \cdots+k_r=0}\,
\det_{1\le s,t\le r}
\ll x_t^{rk_s+s-t}q^{k_s(\frac12 r k_s + s)} \rr
=(q;q)^{r-1}_{\infty}
\prod_{1\le i<j\le r} 
\ll \frac{x_i}{x_j},q\frac{x_j}{x_i}; q \rr_\infty
\end{equation}

\noindent Applying this, with $x_t \to q^{\mu_t-t}$, and $q \to q^{r+d}$, to
\eqref{Eq:CPPgenrlim} yields

\begin{equation}
\label{Eq:CPPprodr}
\CPgfX{r}{\Lambda}
=\frac{(q^{r+d};q^{r+d})^{r-1}_\infty}{(q;q)^r_\infty}
\prod_{1\le i<j\le r}
\ll q^{\mu_i-\mu_j-i+j}, q^{r+d-\mu_i+\mu_j+i-j}; q^{r+d} \rr_{\infty}
\end{equation}

\noindent Theorem {\bf \ref{CPPprodbin}} follows by considering the terms that remain 
after cancelling each factor in the final product with a factor from $(q;q)^r_\infty$, 
possibly with the help of a diagram as in Figure {\bf \ref{TypicalHookProd}}.

\section{$\slchap{r}$ and $\cW_r$ Characters}
\label{characters}

\noindent {\it 
We give simple proofs of the relationship between the generating functions of $r$-cylindric 
partitions and the principally-specialised characters of $\slchap{r}$ and certain characters
of $\cW_r$.}

\subsection{$\slchap{r}$ characters}
\label{SlrChars}

As detailed in Appendix {\bf \ref{AffChars}}, the Weyl-Kac character formula
\cite{kac.book.1990} enables the character of an integrable $\slchap{r}$-module,
with highest weight $\Lambda\in\wtlatd{r}{d}$, to be written in the form

\begin{equation}\label{Eq:SlChar}
\chi^{\slchap{r}}_\Lambda
=
e^\Lambda\frac{{\mathcal N}^{\slchap{r}}_\Lambda}{{\mathcal N}^{\slchap{r}}_0},
\end{equation}
with ${\mathcal N}^{\slchap{r}}_\Lambda$ given by
\begin{equation}\label{Eq:slNum}
{\mathcal N}^{\slchap{r}}_\Lambda=
\sum_{k_1+ \cdots+k_r=0}
\det_{1\le s,t\le r}
\ll x_{s}^{-(r+d)k_{s}-\mu_t+t+\mu_{s}-s}
q^{ (\mu_t-t)k_{s} +\frac12 (r+d) k_s^2 } \rr,
\end{equation}

\noindent where $\mu=\parop(\Lambda)$, $q=e^{-\delta}$ and $x_i/x_{i+1}=e^{-\alpha_i}$
for $1\le i<r$.
\footnote{\,
This implies that $x_1,x_2,\ldots,x_{r}$ are specified only up to an overall factor.
However, although not immediately obvious, ${\mathcal N}^{\slchap{r}}_\Lambda$, as 
defined by \eqref{Eq:slNum}, is independent of such a factor.
}
Applying the Macdonald identity \eqref{Eq:MacId3} to the $\Lambda=0$ case, 
the denominator of \eqref{Eq:SlChar} can be written as 

\begin{equation}\label{Eq:SlDenId}
{\mathcal N}^{\slchap{r}}_0
=(q;q)^{r-1}_{\infty}
\prod_{1\le i<j\le r} \ll \frac{x_i}{x_j},q\frac{x_j}{x_i}; q \rr_{\infty}
\end{equation}

The numerator ${\mathcal N}^{\slchap{r}}_\Lambda$ of \eqref{Eq:SlChar} cannot be written in product 
form. However, there is a product form for the principal specialisation
$\mathrm{Pr}\;{\mathcal N}^{\slchap{r}}_\Lambda$ of \eqref{Eq:slNum}
in which $e^{-\alpha_i}\to q$, for each simple root $\alpha_i$
\footnote{\, 
This holds for any affine Lie algebra $\lieg$:
See \cite[Proposition 10.10]{kac.book.1990} 
and also \cite{lepowsky.milne.1978}.
}.
This specialisation is obtained by substituting $q\to q^r$ and $x_s\to q^{-s}$, leading to

\begin{equation}\label{Eq:SpecSlNum}
\begin{split}
\mathrm{Pr}\;{\mathcal N}^{\slchap{r}}_\Lambda
&=
\sum_{k_1+ \cdots+k_r=0}
\det_{1\le s,t\le r}
\ll q^{(r+d)sk_{s}+(\mu_t-t-\mu_{s}+s)s
+(\mu_t-t)rk_{s} +\frac12 r(r+d)k_s^2 } \rr 
\\
&=
\sum_{k_1+ \cdots+k_r=0}
\prod_{i=1}^r q^{-(\mu_i-i)i}
\det_{1\le s,t\le r}
\ll q^{(\mu_t-t)(rk_s+s)+
(r+d)sk_{s} +\frac12 r(r+d)k_s^2 } \rr
\end{split}
\end{equation}

\noindent In view of \eqref{Eq:SlDenId}, we also have
\begin{equation}\label{Eq:SpecSlDen}
\mathrm{Pr}\;{\mathcal N}^{\slchap{r}}_0
=(q^r;q^r)^{r-1}_{\infty}
\prod_{1\le i<j\le r} \ll q^{j-i},q^{r-j+i}; q^r \rr_{\infty}
=\frac{(q;q)_\infty^r}
      {(q^r;q^r)_\infty}
\end{equation}

Comparing \eqref{Eq:CPPgenrlim} with the ratio of \eqref{Eq:SpecSlNum} and
\eqref{Eq:SpecSlDen} yields

\begin{equation}\label{Eq:CPPaffine}
\CPgfX{r}{\Lambda}=
\frac{1}{(q^r;q^r)_\infty}\,
\frac{\mathrm{Pr}\;{\mathcal N}^{\slchap{r}}_\Lambda}
     {\mathrm{Pr}\;{\mathcal N}^{\slchap{r}}_0}
\end{equation}

The principally specialised character $\mathrm{Pr}\;\chi^{\slchap{r}}_\Lambda$
is obtained from $\chi^{\slchap{r}}_\Lambda$ by setting $e^{-\alpha_i}\to q$, 
for each simple root $\alpha_i$, and setting $e^\Lambda=1$.
Combining \eqref{Eq:SlChar} and \eqref{Eq:CPPaffine} proves the first equality 
in \eqref{Eq:Chars}.

\subsection{$\cW_r$ characters}
\label{WrChars}

Let $r\le p<p'$, with $p$ and $p'$ coprime,
and let $\xi\in\wtlatd{r}{p-r}$ and $\zeta\in\wtlatd{r}{p'-r}$.
As detailed in Appendix {\bf \ref{WrNotes}}, the expression of
\cite{mizoguchi.1991}--\cite{nakanishi.1990}
for the $\cM^{\, p, \pp}_r$ character of the highest weight $\cW_r$ representation
labelled by the pair $\xi$ and $\zeta$ can be written in the form
\footnote{\label{Virasoro} \, 
$\chi^{2,p,p'}_{[p-2-i,i],[p'-2-j,j]}$ is a Virasoro character,
more usually denoted $\chi^{p,p'}_{i+1,j+1}$.
}

\begin{equation}\label{Eq:WrChar}
\begin{split}
\chi^{r,p,p'}_{\xi,\zeta}
&=
\frac{q^{\Delta^r_{\xi,\zeta}}}{(q;q)^{r-1}_\infty}
\sum_{k_1+ \cdots+k_r=0}
q^{p'\sum_{i=1}^r k_i(\frac12 pk_i-\nu_i+i)}
\det_{1\le s,t\le r}
\ll q^{(\mu_t-t)(pk_s-\nu_s+s+\nu_t-t)} \rr,
\end{split}
\end{equation}

\noindent where the partitions $\mu$ and $\nu$ are defined by
$\mu=\parop(\zeta)$ and $\nu=\parop(\xi)$, and
$\Delta^r_{\xi,\zeta}=
\tfrac1{2pp'}|p'(\xi+\rho)-p(\zeta+\rho)|^2 -\tfrac1{24}(r-1)$.

\subsection{$\cW_r$ characters of type $\chi^{\, r, r, r+d}_{0,\zeta}$}
In this paper, we are primarily interested in the $\cM^{\, r, r+d}_r$
characters of $\cW_r$ for $d>0$.
In this case $\xi=0$ and therefore the characters are 
labelled by a single weight $\zeta\in\wtlatd{r}{d}$.
On comparing $\chi^{\, r, r, r+d}_{0,\zeta}$, given by \eqref{Eq:WrChar},
with \eqref{Eq:CPPgenrlim}, we see that the normalised character 
$\chi^{\cW_r}_\zeta=q^{-\Delta^r_{0,\zeta}}\chi^{\, r, r, r+d}_{0,\zeta}$
satisfies

\begin{equation}\label{Eq:WrtoCP}
\chi^{\cW_r}_\zeta
= {(q;q)_\infty}\,
\CPgfX{r}{\zeta}\:.
\end{equation}

\noindent This proves the equality between the first and
third expressions in \eqref{Eq:Chars}.

\subsection{$\cW_r$ characters of type $\chi^{r,p,p'}_{\xi,\zeta}$}
In \cite{gessel.krattenthaler.1997}, Gessel and Krattenthaler
define the notion of $(\boldalpha,\boldbeta)$-cylindric
partitions of type $\lambda/\mu/d$,
which generalises that defined in Section {\bf \ref{Cyl}}.
This enables an analogue of \eqref{Eq:WrtoCP} to be given for the most
general minimal model $\cW_r$ characters $\chi^{r,p,p'}_{\xi,\zeta}$.

For
$\zeta=\sum_{i=0}^{r-1}m_i\Lambda_i$,
$\xi=\sum_{i=0}^{r-1}n_i\Lambda_i$,
and
$\mu=\parop(\zeta)$, 
let $\CPX{r}{\xi,\zeta}$ be the set of all arrays $\pi$
of the form \eqref{Eq:PlaneArray} with $\lambda=(\infty^r)$
that satisfy
\begin{subequations}\label{Eq:PlaneConsB}
\begin{align}\label{Eq:PlaneRow1B}
\pi_{i,j}&\ge\pi_{i,j+1}
&
(1\le i\le r,\,\mu_i\le j),\\
\label{Eq:PlaneRow2B}
\pi_{i,j}&\ge\pi_{i+1,j}-n_i
&
(1\le i<r\,,\max\{\mu_i,\mu_{i+1}\}\le j),\\
\label{Eq:PlaneCylB}
\pi_{r,j}&\ge\pi_{1,j+d}-n_0
&
(\max\{\mu_r,\mu_1-d\}\le j)
\end{align}
\end{subequations}
instead of \eqref{Eq:PlaneCons}.
In the terminology of \cite{gessel.krattenthaler.1997},
$\CPX{r}{\xi,\zeta}$ is the set of $(\boldzero,\boldbeta)$-cylindric
partitions of type $(\infty)^r/\mu/(p'-r)$,
where $\boldbeta=(-n_1,-n_2, \cdots,-n_{r-1},-n_0)$.
With $\CPgfX{r}{\xi,\zeta}$ the
generating function of $\CPX{r}{\xi,\zeta}$,
comparing \eqref{Eq:WrChar} with the appropriate limit of 
\cite[Theorem 3]{gessel.krattenthaler.1997} yields

\begin{equation}
\label{Eq:WrtoCP2}
\chi^{r,p,p'}_{\xi,\zeta}
 = (q;q)_{\infty} \, q^{\Delta^r_{\xi,\zeta}} \, \CPgfX{r}{\xi,\zeta}\:.
\end{equation}

\subsection{$r$-Burge partitions}
\label{r.burge.partitions}
The cylindric partitions $\CPX{r}{\xi,\zeta}$ are also
$r$-Burge partitions, as defined in \cite{belavin.foda.santachiara}.
To see this, define $\hat\pi_{i,j}=\pi_{i,\mu_i+j}$ for $1\le i\le r$
and $j\ge0$.
Then, for each $i$, the sequence
$\hat\pi_i=(\hat\pi_{i,1},\hat\pi_{i,2},\ldots)$ is a partition.
The definition of $\mu=\parop(\zeta)$ implies that
$\mu_i-\mu_{i+1}=m_i$ for $1\le i<r$.
Therefore,
$\hat\pi_{i,j}-\hat\pi_{i+1,j+m_i}=
\pi_{i,j+\mu_i}-\pi_{i+1,j+\mu_i}\ge -n_i$
for $1\le i<r$
and
$\hat\pi_{r,j}-\hat\pi_{1,j+m_0}=
\pi_{i,j}-\pi_{1,j+d}\ge -n_0$.
After appropriately renaming the parameters and indices, we obtain
precisely the conditions \cite[(1)]{belavin.foda.santachiara}.

\section{Reckoning with the abacus}
\label{abacus}

\noindent {\it
We introduce the $r$-string abaci which are in bijection with $r$-cylindric partitions.
}

\subsection{Introducing the abacus}
An $r$-string abacus consists of $r$ infinitely-long strings. Each string is a one-dimensional 
discrete, regular lattice, with infinitely-many equally spaced sites. On each string, there is 
a semi-infinite set of beads. The beads are allowed to occupy the sites on the string, such that 
only a finite number of beads are to the right of any particular position on the string. The 
semi-infinite set of beads acts as a \emph{Dirac sea} from which excitations emerge in the form 
of beads that move to the right along a string to occupy initially vacant positions
\footnote{\,
The $r$-string abacus is an $r$-fold version of a Maya diagram 
\cite{jimbo.miwa.1983, miwa.jimbo.date.book}.
}.

\subsection{From $r$-cylindric partitions to $r$-string abaci}
$r$-cylindric partitions are conveniently represented using $r$-string abaci. Because 
of level-rank duality
\footnote{\,
See subsection {\bf \ref{level.rank.duality}}.
}, 
two such representations are possible, one with $r$ strings and one with $d$ strings.
For the purposes of the current paper, we will use the representation that has $r$ strings.

The $r$-string abacus that corresponds to an $r$-cylindric partition is obtained by associating 
each vacancy on the $t$-th string of the abacus with an entry in the $t$th row of the cylindric 
partition. Specifically, scanning the $t$-th row the $r$-cylindric partition from left to right,
the $i$-th entry gives the number of beads that appear on the $t$-th string to the right of the 
$i$-th vacancy.

\subsubsection{Example} 
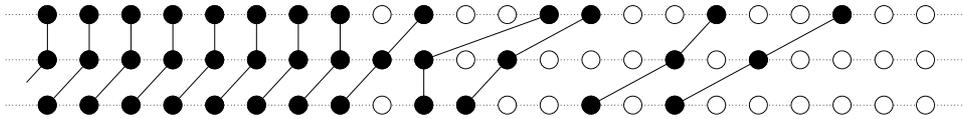
\begin{figure}[ht]
\begin{center}
\psset{yunit=6.0mm,xunit=5.5mm}
\begin{pspicture}(-9,0)(14,4.5)
\psset{linewidth=0.25pt,linestyle=dotted,dotsep=1.0pt,linecolor=black}
\multips(0,1)(0,1){3}{\psline{-}(-9,0)(14,0)}
\psset{dotsize=2.5mm,fillstyle=solid,fillcolor=black,dotstyle=o}
\psset{linewidth=0.35pt,linestyle=solid,linecolor=black}
\multips(-2,1)(-1,0){7}{
  \psset{fillstyle=none}\psline{-}(0,0)(1,1)(1,2) 
  \psset{fillstyle=solid}\psdots(0,0)(1,1)(1,2)   
}
\multips(-9,1)(-1,0){1}{
  \psset{fillstyle=none}\psline{-}(0.5,0.5)(1,1)(1,2) 
  \psset{fillstyle=solid}\psdots(1,1)(1,2)   
}
\multips(-1,1)(-1,0){1}{
  \psset{fillstyle=none}\psline{-}(0,0)(1,1)(2,2) 
  \psset{fillstyle=solid}\psdots(0,0)(1,1)(2,2)   
}
\multips(1,1)(-1,0){1}{
  \psset{fillstyle=none}\psline{-}(0,0)(0,1)(3,2) 
  \psset{fillstyle=solid}\psdots(0,0)(0,1)(3,2)   
}
\multips(2,1)(-1,0){1}{
  \psset{fillstyle=none}\psline{-}(0,0)(1,1)(3,2) 
  \psset{fillstyle=solid}\psdots(0,0)(1,1)(3,2)   
}
\multips(5,1)(-2,0){1}{
  \psset{fillstyle=none}\psline{-}(0,0)(2,1)(3,2) 
  \psset{fillstyle=solid}\psdots(0,0)(2,1)(3,2)   
}
\multips(7,1)(-2,0){1}{
  \psset{fillstyle=none}\psline{-}(0,0)(2,1)(4,2) 
  \psset{fillstyle=solid}\psdots(0,0)(2,1)(4,2)   
}
\psset{dotsize=2.5mm,fillstyle=none,fillcolor=white,dotstyle=o}
\psdots(0,1)(3,1)(4,1)(6,1)(8,1)(9,1)(10,1)(11,1)(12,1)(13,1)
\psdots(2,2)(4,2)(5,2)(6,2)(8,2)(10,2)(11,2)(12,2)(13,2)
\psdots(0,3)(2,3)(3,3)(6,3)(7,3)(9,3)(10,3)(12,3)(13,3)
\end{pspicture}
\end{center}
\caption{\it 
Typical $r=3$, $d=4$ abacus, for $\Lambda=3\Lambda_0+\Lambda_2$.
The beads are the black circles. The vacancies are the white circles.
The yokes are the 2-segment strings that connect sets of 3 beads on
different strings without intersecting.
}
\label{TypicalAbacusr=3}
\end{figure}

The $r=3$, $d=4$ cylindric partition of \eqref{Eq:TypCPP1} corresponds to the 3-string abacus 
in Figure {\bf \ref{TypicalAbacusr=3}}. In this example, the beads have been yoked together, 
that is, attached with finite-length strings that consist of $(r-1)$ segments each, in such 
a way that they allow for sets of vacancies. Far to the left, the yoked $r$-tuples 
of beads are all of the same shape. This shape corresponds to the partition $\parop(\Lambda)$
in the definition of the $r$-cylindric partition. These yoked $r$-tuples are vertical 
when $\Lambda=d\Lambda_0$. Each shape of yokes, and each set of vacancies corresponds 
to a unique $r$-cylindric partition.

\subsection{The formation of an $r$-cylindric partition} 
For $j>0$, let $\delta_j$ be the number of entries $j$ in ($r$ consecutive rows of)
the cylindric partition $\pi\in\CPX{r}{\Lambda}$.
Then $\delta_j$ is the number of vacancies between the $j$-th and $(j+1)$-th yoked 
$r$-tuples of the corresponding abacus, counting the $r$-tuples from the right.
We immediately have

\begin{equation}\label{Eq:piWtDef}
|\pi|=\sum_{j=1}^\infty j\delta_j
\end{equation}

\noindent We refer to the sequence $\bolddelta(\pi)=(\cdots, \delta_3, \delta_2, \delta_1)$
as the \emph{formation} of $\pi$. Note that if $\pi\in\CPX{r}{\Lambda}(a)$ then $\delta_j=0$ 
for $j>a$.

\subsection{The cylindric abacus and $\Lambda$ yokes}
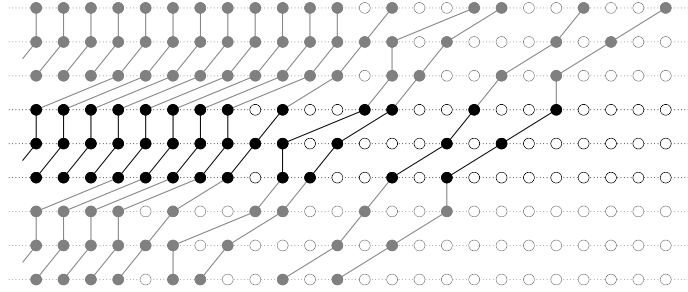
\begin{figure}[ht]
\begin{center}
\psset{yunit=4.5mm,xunit=3.6mm}
\begin{pspicture}(-9,-3)(15,7.5)
\psset{linewidth=0.25pt,linestyle=dotted,dotsep=1.0pt,linecolor=black}
\multips(0,1)(0,1){3}{\psline{-}(-9,0)(16,0)}
\psset{linewidth=0.25pt,linestyle=dotted,dotsep=1.0pt,linecolor=gray}
\multips(0,0)(0,-1){3}{\psline{-}(-9,0)(16,0)}
\multips(0,4)(0,1){3}{\psline{-}(-9,0)(16,0)}
\psset{linewidth=0.35pt,linestyle=solid,linecolor=gray}
\multips(-1,3)(-1,0){8}{\psline{-}(0,0)(3,1)}
\multips(-5,0)(-1,0){4}{\psline{-}(0,0)(3,1)}
\psline{-}(1,3)(3,4)\psline{-}(4,3)(5,4)\psline{-}(5,3)(6,4)
\psline{-}(8,3)(9,4)\psline{-}(11,3)(11,4)
\psline{-}(-3,0)(-1,1)\psline{-}(0,0)(1,1)\psline{-}(1,0)(2,1)
\psline{-}(4,0)(5,1)\psline{-}(7,0)(7,1)
\psset{dotsize=1.5mm,fillstyle=solid,fillcolor=black,dotstyle=o}
\psset{linewidth=0.35pt,linestyle=solid,linecolor=black}
\multips(-2,1)(-1,0){7}{
  \psset{fillstyle=none}\psline{-}(0,0)(1,1)(1,2) 
  \psset{fillstyle=solid}\psdots(0,0)(1,1)(1,2)   
}
\multips(-9,1)(-1,0){1}{
  \psset{fillstyle=none}\psline{-}(0.5,0.5)(1,1)(1,2) 
  \psset{fillstyle=solid}\psdots(1,1)(1,2)   
}
\multips(-1,1)(-1,0){1}{
  \psset{fillstyle=none}\psline{-}(0,0)(1,1)(2,2) 
  \psset{fillstyle=solid}\psdots(0,0)(1,1)(2,2)   
}
\multips(1,1)(-1,0){1}{
  \psset{fillstyle=none}\psline{-}(0,0)(0,1)(3,2) 
  \psset{fillstyle=solid}\psdots(0,0)(0,1)(3,2)   
}
\multips(2,1)(-1,0){1}{
  \psset{fillstyle=none}\psline{-}(0,0)(1,1)(3,2) 
  \psset{fillstyle=solid}\psdots(0,0)(1,1)(3,2)   
}
\multips(5,1)(-2,0){1}{
  \psset{fillstyle=none}\psline{-}(0,0)(2,1)(3,2) 
  \psset{fillstyle=solid}\psdots(0,0)(2,1)(3,2)   
}
\multips(7,1)(-2,0){1}{
  \psset{fillstyle=none}\psline{-}(0,0)(2,1)(4,2) 
  \psset{fillstyle=solid}\psdots(0,0)(2,1)(4,2)   
}
\psset{dotsize=1.5mm,fillstyle=none,fillcolor=white,dotstyle=o}
\psdots(0,1)(3,1)(4,1)(6,1)(8,1)(9,1)(10,1)(11,1)
\psdots(2,2)(4,2)(5,2)(6,2)(8,2)(10,2)(11,2)
\psdots(0,3)(2,3)(3,3)(6,3)(7,3)(9,3)(10,3)
\multips(12,1)(1,0){4}{\psdots(0,0)(0,1)(0,2)}
\psset{dotsize=1.5mm,fillstyle=solid,fillcolor=gray,dotstyle=o}
\psset{linewidth=0.35pt,linestyle=solid,linecolor=gray}
\multips(2,4)(-1,0){11}{
  \psset{fillstyle=none}\psline{-}(0,0)(1,1)(1,2) 
  \psset{fillstyle=solid}\psdots(0,0)(1,1)(1,2)   
}
\multips(-9,4)(-1,0){1}{
  \psset{fillstyle=none}\psline{-}(0.5,0.5)(1,1)(1,2) 
  \psset{fillstyle=solid}\psdots(1,1)(1,2)   
}
\multips(3,4)(-1,0){1}{
  \psset{fillstyle=none}\psline{-}(0,0)(1,1)(2,2) 
  \psset{fillstyle=solid}\psdots(0,0)(1,1)(2,2)   
}
\multips(5,4)(-1,0){1}{
  \psset{fillstyle=none}\psline{-}(0,0)(0,1)(3,2) 
  \psset{fillstyle=solid}\psdots(0,0)(0,1)(3,2)   
}
\multips(6,4)(-1,0){1}{
  \psset{fillstyle=none}\psline{-}(0,0)(1,1)(3,2) 
  \psset{fillstyle=solid}\psdots(0,0)(1,1)(3,2)   
}
\multips(9,4)(-2,0){1}{
  \psset{fillstyle=none}\psline{-}(0,0)(2,1)(3,2) 
  \psset{fillstyle=solid}\psdots(0,0)(2,1)(3,2)   
}
\multips(11,4)(-2,0){1}{
  \psset{fillstyle=none}\psline{-}(0,0)(2,1)(4,2) 
  \psset{fillstyle=solid}\psdots(0,0)(2,1)(4,2)   
}
\multips(-6,-2)(-1,0){3}{
  \psset{fillstyle=none}\psline{-}(0,0)(1,1)(1,2) 
  \psset{fillstyle=solid}\psdots(0,0)(1,1)(1,2)   
}
\multips(-9,-2)(-1,0){1}{
  \psset{fillstyle=none}\psline{-}(0.5,0.5)(1,1)(1,2) 
  \psset{fillstyle=solid}\psdots(1,1)(1,2)   
}
\multips(-5,-2)(-1,0){1}{
  \psset{fillstyle=none}\psline{-}(0,0)(1,1)(2,2) 
  \psset{fillstyle=solid}\psdots(0,0)(1,1)(2,2)   
}
\multips(-3,-2)(-1,0){1}{
  \psset{fillstyle=none}\psline{-}(0,0)(0,1)(3,2) 
  \psset{fillstyle=solid}\psdots(0,0)(0,1)(3,2)   
}
\multips(-2,-2)(-1,0){1}{
  \psset{fillstyle=none}\psline{-}(0,0)(1,1)(3,2) 
  \psset{fillstyle=solid}\psdots(0,0)(1,1)(3,2)   
}
\multips(1,-2)(-2,0){1}{
  \psset{fillstyle=none}\psline{-}(0,0)(2,1)(3,2) 
  \psset{fillstyle=solid}\psdots(0,0)(2,1)(3,2)   
}
\multips(3,-2)(-2,0){1}{
  \psset{fillstyle=none}\psline{-}(0,0)(2,1)(4,2) 
  \psset{fillstyle=solid}\psdots(0,0)(2,1)(4,2)   
}
\psset{dotsize=1.5mm,fillstyle=none,fillcolor=white,dotstyle=o,linecolor=gray}
\psdots(4,4)(7,4)(8,4)(10,4)(12,4)(13,4)(14,4)(15,4)
\psdots(6,5)(8,5)(9,5)(10,5)(12,5)(14,5)(15,5)
\psdots(4,6)(6,6)(7,6)(10,6)(11,6)(13,6)(14,6)
\psdots(-4,-2)(-1,-2)(0,-2)(2,-2)(4,-2)(5,-2)(6,-2)(7,-2)
\psdots(-2,-1)(0,-1)(1,-1)(2,-1)(4,-1)(6,-1)(7,-1)
\psdots(-4,0)(-2,0)(-1,0)(2,0)(3,0)(5,0)(6,0)
\multips(8,-2)(1,0){8}{\psdots(0,0)(0,1)(0,2)}
\end{pspicture}
\end{center}
\caption{\it 
Extension of $r=3$, $d=4$ abacus from Figure {\bf \ref{TypicalAbacusr=3}}
}
\label{TypicalAbacusr=3B}
\end{figure}

Corresponding to the augmented viewpoint \eqref{Eq:TypCPP2} of the cylindric partition,
we can place infinitely-many copies of the $r$ strings of the abacus above and below 
the original abacus, with each copy displaced by $d$ positions further to the right with
respect to the copy immediately below it. Equivalently, the $r$ strings may be placed on 
a cylinder. Doing this in the case of Figure {\bf \ref{TypicalAbacusr=3}} leads to a diagram, 
a portion of which is shown, in Figure {\bf \ref{TypicalAbacusr=3B}}.

In view of the above cylindrical extension, each yoked $r$-tuple of beads should be 
considered as being of infinite length, consisting of the same repeating sequence of 
$r$ segments whose horizontal lengths, $m_0,m_1, \cdots,m_{r-1}$, sum to $d$. This 
way, each $r$-tuple is associated with a dominant affine weight
$\Lambda=[m_0,m_1, \cdots,m_{r-1}]\in\wtlatd{r}{d}$.
We call this a $\Lambda$-yoke.

\subsection{$\Lambda$ yokels}
For $\Lambda\in\wtlatd{r}{d}$, we define a $\Lambda$-yokel to be a
semi-infinite sequence
$\boldLambda=(\cdots,\Lambda^{(3)},\Lambda^{(2)},\Lambda^{(1)})$
of elements $\Lambda^{(j)}\in\wtlatd{r}{d}$ for $j\ge1$,
for which there exists $L$ such that $\Lambda^{(j)}=\Lambda$
for all $j>L$.
Let $\ell(\boldLambda)$ be the smallest such $L$.
Note that if $\pi\in\CPX{r}{\Lambda}(a)$,
then $\ell(\boldLambda(\pi))\le a$.
For $\pi\in\CPX{r}{\Lambda}$, we define the $\Lambda$-yokel
$\boldLambda(\pi)=(\cdots,\Lambda^{(3)},\Lambda^{(2)},\Lambda^{(1)})$,
by setting $\Lambda^{(j)}$ to be such that,
counting from the right,
the $j$th yoke of the abacus corresponding to $\pi$
is a $\Lambda^{(j)}$-yoke.

For each $\pi$, we can thus use the abacus to obtain a pair
$(\boldLambda(\pi),\bolddelta(\pi))$, from which
$\pi$ can be uniquely reconstructed.
However, not every pair $(\boldLambda,\bolddelta)$
in which $\boldLambda$ is a $\Lambda$-yokel
and $\bolddelta=(\cdots,\delta_3,\delta_2,\delta_1)$
is an arbitrary sequence of non-negative integers,
arises in this way.

\subsection{Yokes apart}
\label{Yokel}

The process of moving a yoke one position horizontally to the left,
without touching another yoke is known as a \emph{lift}
\footnote{\, 
In this formulation, the term {\it \lq lift\rq\,} is a misnomer, and it makes more 
sense to use the {\it \lq patterns\rq\,} of \cite{jimbo.misra.miwa.okado.1991}.
We explain the correspondence between these patterns and cylindric partitions in 
Appendix {\bf \ref{Kyoto}}. The notions of {\it \lq tightening\rq\,} and {\it \lq tight\rq\,} 
in \cite{tingley.2008} are equivalent to those of {\it \lq lift\rq\,} and {\it \lq highest-lift\rq\,} 
here.
}.
In the example of Figure {\bf \ref{TypicalAbacusr=3B}}, we see that lifts 
can be performed on both the first and second yokes. Correspondingly, 
an \emph{inverse lift} is a moving of a yoke to the right.

For $\Lambda', \Lambda'' \in \wtlatd{r}{d}$, consider a $\Lambda'$-yoke
on an abacus to the left of a $\Lambda''$-yoke, with $\delta$ the number 
of vacancies between them.  We see that performing a lift on the 
$\Lambda''$-yoke, or an inverse lift on the $\Lambda'$-yoke,
results in the two yokes being separated by $\delta'=\delta-r$ vacancies.
By repeating the process, we see that there is a minimal possible
number of vacancies between a $\Lambda'$-yoke and a $\Lambda''$-yoke,
which we denote $\Delta(\Lambda', \Lambda'')$, and that

\begin{equation}
\label{Eq:rDist}
\delta\in r\ZZp+\Delta(\Lambda',\Lambda'')
\end{equation}

\subsubsection{Examples}
We find that

\begin{equation}
\Delta \ll [4,0,0],[3,1,0] \rr =1, \quad
\Delta \ll [4,0,0],[0,3,1] \rr =5, \quad
\Delta \ll [4,0,0],[1,0,3] \rr =6
\end{equation}

\noindent Note that it is not necessarily the case that $\Delta(\Lambda',\Lambda'')<r$.
In the $r=2$ case, we obtain the explicit expression

\begin{equation}\label{Eq:Distsr=2a}
\Delta \ll [d-x,x],[d-y,y] \rr =|x-y|,
\end{equation}

\noindent which results from direct consideration of suitable yokes on the abacus.

\subsection{Highest-lift abaci}

Let $\pi\in\CPX{r}{\Lambda}$ and let
$\boldLambda(\pi)=(\cdots,\Lambda^{(3)},\Lambda^{(2)},\Lambda^{(1)})$
and $\bolddelta(\pi)=(\cdots,\delta_3,$ $\delta_2$,$ \delta_1)$.
Let $\pi'$ be obtained from $\pi$ by applying an inverse lift
to the $j$th yoke.
If $\bolddelta(\pi')=(\cdots,\delta'_3,\delta'_2,\delta'_1)$
then $\delta'_j=\delta_j+r$,
$\delta'_{j-1}=\delta_{j-1}-r$ (if $j>1$)
and $\delta'_k=\delta_k$ otherwise.
Then \eqref{Eq:piWtDef} gives $|\pi'|=|\pi|+r$.
It follows that the generating function of all abaci obtained by
applying inverse lifts to $\pi$ is $q^{|\pi|}(q^r;q^r)_\infty^{-1}$.

A cylindric partition $\pi$ (or abacus) for which no lifts are
possible is known as a \emph{highest-lift} configuration.
Let $\HLX{r}{\Lambda}\subset\CPX{r}{\Lambda}$ be the subset
of highest-lift configurations.
Note that $\pi\in\HLX{r}{\Lambda}$ is completely characterised
by $\boldLambda(\pi)$.
Moreover, every $\pi'\in\CPX{r}{\Lambda}$ for which
$\boldLambda(\pi')=\boldLambda(\pi)$ is obtained from $\pi$
by a sequence of inverse lifts.
It follows that if $\HLgfX{r}{\Lambda}$
denotes the generating function for $\HLX{r}{\Lambda}$ then

\begin{equation}\label{Eq:CPinH}
\CPgfX{r}{\Lambda}
=\frac{1}{(q^r;q^r)_\infty}\,\HLgfX{r}{\Lambda}
\end{equation}

\noindent Comparing this result with the first equality in \eqref{Eq:Chars},
which was proved in section {\bf \ref{SlrChars}}, then shows that

\begin{equation}\label{Eq:HisChar}
\HLgfX{r}{\Lambda}=\text{Pr}\,\chi^{\slchap{r}}_\Lambda
\end{equation}

\subsection{Note on crystal graphs}
The previous result is also a consequence of the crystal graph theory of representations 
of $\slchap{r}$ described in \cite{jimbo.misra.miwa.okado.1991}. There, so-called `patterns' 
are used to label the nodes of the crystal graph.
As we show in Appendix {\bf \ref{Kyoto}}, these patterns encode cylindric partitions.
The subset of the full set of patterns which, in \cite{jimbo.misra.miwa.okado.1991},
are known as `highest-lift' are in bijection with the nodes of the
crystal graph of an irreducible representation of $\slchap{r}$.
These highest-lift patterns correspond to those cylindric partitions
that are designated highest-lift in this section.
Note, however, that the crystal graph theory
provides a combinatorial model for the full characters 
$\chi^{\slchap{r}}_\Lambda$ of $\slchap{r}$.

\section{The Andrews-Gordon-Bressoud identities. Outline of results}
\label{andrews.gordon.bressoud}

\noindent {\it
Our main results are Theorem {\bf \ref{CPAG}} and Corollary {\bf \ref{Cor:CPAG}}
}

\subsection{Main results}

In this and the following sections, we concentrate on the $r=2$ case.
Here $\Lambda\in\wtlatd{2}{d}$ takes the form $[d-i,i]$ for $0\le i\le d$,
and $\parop(\Lambda)=(i,0)$.
We set $p\!=\!d+2$.
Then, expanding the determinant of \eqref{Eq:CPPgenr} yields

\begin{equation}\label{Eq:CPPbosr=2f}
\begin{split}
\CPgfX{2}{\Lambda}(a)&=
\sum_{k_1\in\ZZ}
\frac{q^{k_1(2k_1-1)p+2k_1(i+1)}}{(q;q)_{a+2k_1}(q;q)_{a-2k_1}}
-\sum_{k_1\in\ZZ}
\frac{q^{k_1(2k_1-1)p-(2k_1-1)(i+1)}}{(q;q)_{a+2k_1-1}(q;q)_{a-2k_1+1}}\\
&=
\sum_{j\in\ZZ}
(-1)^j\frac{q^{\frac12j(j+1)p-j(i+1)}}{(q;q)_{a-j}(q;q)_{a+j}}\,,
\end{split}
\end{equation}

\noindent having combined the two summations, using $j=-2k_1$ in the first 
and $j=2k_1-1$ in the second. Taking the $a\to\infty$ limit then gives

\begin{equation}\label{Eq:CPPbosr=2}
\CPgfX{2}{\Lambda}
=\frac{1}{(q;q)^2_\infty}
\sum_{j\in\ZZ}
(-1)^j{q^{\frac12j(j+1)p-j(i+1)}}
\end{equation}

\noindent From \eqref{Eq:CPPprodr}, we also have

\begin{equation}\label{Eq:CPPprodr=2}
\CPgfX{2}{\Lambda}
=\frac{(q^{i+1},q^{p-i-1},q^p;q^p)_\infty}{(q;q)^2_\infty}
\end{equation}

\noindent When $p$ is odd, as indicated in footnote {\bf \ref{Virasoro}},
the Virasoro character $\chi^{2,p}_{1,i+1}=(q;q)_\infty\CPgfX{2}{\Lambda}$.

\subsection{Bailey chains}
Expression \eqref{Eq:CPPbosr=2f} is at the heart of Bailey chain methods to derive 
generalisations of the Rogers-Ramanu\-jan identities \cite{andrews.1984}--\cite{foda.quano.02}
\footnote{\,
For a review and further reference to Bailey chain methods, see \cite{warnaar.2001}.
}.
These methods lead to non-negative sum expressions, which when, in the case of 
\eqref{Eq:CPPbosr=2f}, are equated with \eqref{Eq:CPPprodr=2}, yield generalisations of the 
Rogers-Ramanu\-jan identities that are originally due to Andrews \cite{andrews.1974}, Gordon 
\cite{gordon.1961} and Bressoud \cite{bressoud.1980}.
Here, we make a modest addition to the understanding of these
identities, by showing how they arise from simple manipulations
of cylindric partitions.
In fact, the manipulations will be made on what we refer to as
{\it decorated $\B$-paths},
which are modified versions of Bressoud's lattice paths \cite{bressoud.1989}.
They come in two versions, appropriate to the cases of odd and even $d$,
and are defined in Sections {\bf \ref{DBpaths}} and {\bf \ref{EBpaths}} below.
The bijections between such paths and the cylindric partitions
are described in Sections {\bf \ref{BijDBpaths}} and {\bf \ref{BijDEBpaths}}.

For convenience, we will state the sum-type expressions for
$\CPgfX{2}{\Lambda}(a)$ and $\CPgfX{2}{\Lambda}$ in the form of a theorem
and a corollary.
Note that there are different forms for the cases of odd and even $d$
These correspond to the coprime and non-coprime cases discussed above.
The theorem makes use of generalised $q$-multinomials
$\qbinom{a}{\lambda}^{(f)}$ expressed in terms of partitions.
For $a,f\ge0$ and $\lambda\in\partset{k}$, define

\begin{equation}
\qbinom{a}{\lambda}^{(f)}=
\begin{cases}
\dfrac{(q^f;q^f)_a}
     {(q;q)_{a-\lambda_1}(q;q)_{\lambda_1-\lambda_2}(q;q)_{\lambda_2-\lambda_3}
     \cdots(q;q)_{\lambda_{k-1}-\lambda_k}(q^f;q^f)_{\lambda_k}}
     &\text{if $|\lambda_1|\le a$,}\\
     0
     &\text{if $|\lambda_1|>a$}\\
\end{cases}
\end{equation}

\noindent If $f=1$, then the superscript `$(1)$' can be omitted. Note that this 
generalises the definition of the $q$-binomial because for the one part partition 
$\lambda=(n)$ we have $\qbinom{a}{\lambda}=\qbinom{a}{n}$.

\begin{theorem}
\label{CPAG}
Let $d\ge2$ and $0\le i\le k$ with $k=\lfloor d/2\rfloor$.
Then, for $\Lambda=[d-i,i]\in\wtlatd{2}{d}$ and $a\ge0$,
\begin{equation}\label{Eq:CPAG}
\CPgfX{2}{\Lambda}(a)
=
\begin{cases}
\displaystyle
\frac1{(q;q)_a}
\sum_{\lambda\in\partset{k}}
q^{\lambda_1^2+\lambda_2^2+ \cdots+\lambda_k^2
      +\lambda_{i+1}+\lambda_{i+2}+ \cdots+\lambda_k}
\qbinom{a}{\lambda}
&\text{if $d$ is odd,}\\[5mm]
\displaystyle
\frac1{(q^2;q^2)_a}
\sum_{\lambda\in\partset{k}}
q^{\lambda_1^2+\lambda_2^2+ \cdots+\lambda_k^2
      +\lambda_{i+1}+\lambda_{i+2}+ \cdots+\lambda_k}
\qbinom{a}{\lambda}^{(2)}
&\text{if $d$ is even.}
\end{cases}
\end{equation}
\end{theorem}

\noindent Taking the limit $a\to\infty$ of \eqref{Eq:CPAG} leads to the following:

\begin{corollary}
\label{Cor:CPAG}
Let $d\ge2$ and $0\le i\le k$ with $k=\lfloor d/2\rfloor$.
Then, for $\Lambda=[d-i,i]\in\wtlatd{2}{d}$,
\begin{equation}\label{Eq:CPAGc}
\CPgfX{2}{\Lambda}
=
\begin{cases}
\displaystyle
\frac1{(q;q)_\infty}
\sum_{\lambda\in\partset{k}}
\frac{q^{\lambda_1^2+\lambda_2^2+ \cdots+\lambda_k^2
      +\lambda_{i+1}+\lambda_{i+2}+ \cdots+\lambda_k}}
     {(q;q)_{\lambda_1-\lambda_2}(q;q)_{\lambda_2-\lambda_3}
     \cdots(q;q)_{\lambda_{k-1}-\lambda_k}(q;q)_{\lambda_k}}
&\text{if $d$ is odd,}\\[5mm]
\displaystyle
\frac1{(q;q)_\infty}
\sum_{\lambda\in\partset{k}}
\frac{q^{\lambda_1^2+\lambda_2^2+ \cdots+\lambda_k^2
      +\lambda_{i+1}+\lambda_{i+2}+ \cdots+\lambda_k}}
     {(q;q)_{\lambda_1-\lambda_2}(q;q)_{\lambda_2-\lambda_3}
     \cdots(q;q)_{\lambda_{k-1}-\lambda_k}(q^2;q^2)_{\lambda_k}}
&\text{if $d$ is even.}
\end{cases}
\end{equation}
\end{corollary}

\noindent Equating these expressions \eqref{Eq:CPAGc} with \eqref{Eq:CPPprodr=2} 
then yields the aforementioned Andrews-Gordon-Bressoud Rogers-Ramanu\-jan type identities.

\subsection{Plan of proof}
To prove Theorem {\bf \ref{CPAG}}, we first show that the RHS of {\bf \ref{CPAG}} 
gives the generating functions for the corresponding decorated $\B$-paths.
This is done in Theorems {\bf \ref{Achain}} and {\bf \ref{Bchain}}
for the cases of odd and even $d$ respectively.
The proof is then completed by showing that there are weight-preserving
bijections between the cylindric partitions and the corresponding
decorated $\B$-paths.
Lemmas {\bf \ref{WtPresOdd}} and {\bf \ref{WtPresEven}} state the implied 
identities between their generating functions.

\section{$\B$-paths, decorated paths and transforms}
\label{paths.decorated.paths.and.transforms}

\noindent{\it
We recall Bressoud's paths, define a decorated version of these paths, then recall 
Bressoud's transforms that interpolate paths that satisfy different conditions.
}

\subsection{$\B$-paths}

We define a $\B$-path $\hbress$ to be a semi-infinite sequence $\hbress = (h_0, h_1, h_2$, $\cdots)$ 
satisfying $h_i\in\ZZp$ and $h_{i+1}-h_{i}=\{0,\pm1\}$ for $i\ge0$, with $h_{i+1}=h_{i}$ only if $h_i=0$
\footnote{\,
The $\B$ in $\B$-paths, {\it etc.} is for Bressoud who introduced these very paths, as well 
as the transforms that relate paths that satisfy different conditions in \cite{bressoud.1989}.
Since the work in \cite{bressoud.1989} was motivated by Burge's work \cite{burge.1993}, which 
in turn was motivated by Bailey's work \cite{bailey.01, bailey.02}, $\B$ may equally stand for 
Bressoud, Burge and Bailey. Another version of these paths was used by Warnaar in 
\cite{warnaar.01, warnaar.02} to obtain non-negative sum forms of characters in minimal 
Virasoro models labelled by $p$ and $\pp = p+1$.
{\it \lq Half-lattice paths\rq\,} were defined in \cite{bfournier.mathieu.welsh.01, bfournier.mathieu.welsh.02}, 
and used to derive non-negative $q$-series expressions for Virasoro characters related 
to non-thermal pertubations of minimal models. 
{\it \lq Fused paths\rq\,} were defined in \cite{tartaglia.pearce}, 
and used to study alternating-sign $q$-series expressions for Virasoro characters in higher-level 
non-unitary minimal models.
}
For $0\le b\le k$ define $\Adb$ to be the set of all $\B$-paths $\hbress=(h_0, h_1, h_2$, $\cdots)$ 
for which $h_0=b$ and $0\le h_i\le k$ for $i>0$, and for which there exists $L\ge0$ with $\hbress_i=0$ 
for all $i\ge L$. Such paths were introduced in \cite{bressoud.1989}
\footnote{\,
In \cite{foda.lee.pugai.welsh.1998, foda.welsh.2000}, an equivalent version of Bressoud's $\B$-paths 
was introduced. In the latter version, the paths must always change heights.
}.

\subsubsection{Example}
A \emph{picture} of $\hbress\in\Adb$ is obtained by linking the points $(0,\hbress_0)$, $(1,\hbress_{1})$,
$(2,\hbress_2), (3,\hbress_3)$ on the plane. Figure {\bf \ref{TypicalA11a}} shows the picture of 
a typical element $\hbress\in\Adset{5}{3}$.

\begin{figure}[ht]
\begin{center}
\psset{yunit=0.32cm,xunit=0.28cm}
\begin{pspicture}(-2,-1)(32,6.5)
\rput[bl](0,0){
\psset{linewidth=0.25pt,linestyle=dashed, dash=2.5pt 1.5pt,linecolor=gray}
\multips(0,1)(0,1){5}{\psline{-}(0,0)(32,0)}
\multips(0,0)(0,0){5}{\psline{-}(0,0)(32,0)}
\multips(1,0)(1,0){31}{\psline{-}(0,0)(0,5)}
\multips(0,0)(0,0){31}{\psline{-}(0,0)(0,5)}
\rput[r](-0.5,5){\scriptsize $5$}
\rput[r](-0.5,4){\scriptsize $4$}
\rput[r](-0.5,3){\scriptsize $3$}
\rput[r](-0.5,2){\scriptsize $2$}
\rput[r](-0.5,1){\scriptsize $1$}
\rput[r](-0.5,0){\scriptsize $0$}
\psset{linewidth=0.7pt,fillstyle=none,linestyle=solid,linecolor=black}
\psline(0,3)(2,1)(4,3)(7,0)(8,0)(9,1)(10,0)(15,5)(17,3)(18,4)(22,0)(24,0)
       (26,2)(28,0)(32,0)
}
\end{pspicture}
\end{center}
\caption{\it 
Typical $\hbress\in\Adset{5}{3}$
}
\label{TypicalA11a}
\end{figure}

\subsubsection{Peaks and local weights}
A path $\hbress$ has a \emph{peak} at $i > 0$, if both $\hh_{i-1}<\hh_i$ and $\hh_{i+1}<h_i$. 
Each peak is assigned a weight equal to its position $i$. These weights are the local weights 
referred to in subsection \ref{one.dimensional.paths.on.restricted.one.dimensional.lattices}.

\subsubsection{The weight of a path}
The total weight, or simply the weight $\wt(\hbress)$ of a path $\hbress\in\Adb$ is the sum 
of its local weights, that is the sum of the positions of its peaks,

\begin{equation}
\label{Eq:BressWtDef}
\wt(\hbress)=
\sum_{\substack{i>0\\ \hh_{i-1}<h_{i}>\hh_{i+1}}} i
\end{equation}

\noindent The generating function of the weighted paths $\AdbGF$ is

\begin{equation}
\label{Eq:AdbGF}
\AdbGF = \sum_{\hbress\in\Adb} q^{\wt(\hbress)}
\end{equation}

\subsubsection{The number of peaks of a path}
An important attribute of a $\B$-path $\hbress$ is the number of peaks $\npt(\hbress)$ of the path. 
\footnote{\,
The number of peaks of a $\B$-path is called the {\it \lq number of particles\rq\,} in 
physics-motivated literature, including \cite{bosonic.fermionic.01}--\cite{bosonic.fermionic.06}, 
and related works.
}.
We define $\Adb(a)=\{\hbress\in\Adb\,|\,\npt(\hbress)=a\}$ to be the set of paths with the same number 
of peaks $a$. The corresponding generating function is

\begin{equation}
\label{Eq:AdbGF.a}
\AdbGF(a)
= \sum_{\hbress\in\Adb(a)} q^{\wt(\hbress)}
\end{equation}

\begin{lemma}\label{U1}
For $k>0$,
\begin{equation}\label{Eq:U1}
\AGF{k}{k}(a)=q^a\,\AGF{k}{k-1}(a)
\end{equation}
\end{lemma}

\Proof Each $\hbress\in\Adset{k}{k}(a)$ is necessarily such that
$\hh_0=k$ and $\hh_1=k-1$.
Therefore $i=1$ not a peak.
Removing the first segment of $\hbress$ therefore results in
an element $\hbress'\in\Adset{k}{k-1}(a)$.
Moreover, every element $\hbress'\in\Adset{k}{k-1}(a)$ arises
from a unique $\hbress\in\Adset{k}{k}(a)$ in this way.
Because each peak of $\hbress'$ is one less than the corresponding peak
in $\hbress$, and there are $a$ peaks, \eqref{Eq:U1} immediately follows.
\cqfd

\subsection{Decorated $\B$-paths}
\label{DBpaths}

A decorated $\B$-path $\thbress$ is a pair $(\hbress,\ho)$ where $\hbress$ 
is a $\B$-path and $\ho$ is any sequence $\ho=(\ho_0,\ho_1,\ho_2, \cdots)$
of non-negative integers, for which there exists $M\ge0$ such that $\ho_i=0$ 
for $i>M$. 

\subsubsection{Deaks}
We say that $\thbress$ has $\ho_i$ \emph{deaks}, for `degenerate peaks', at 
$i$, these being in addition to a normal peak if $\hh_{i-1}<\hh_i>\hh_{i+1}$.
For $0\le b\le k$ we define $\tAdb$ to be the set of all decorated $\B$-paths 
$\thbress=(\hbress,\ho)$ for which $\hbress\in\Adb$. The picture of 
$(\hbress,\ho)\in\tAdb$ is obtained from that of $\hbress$ by, for each $i$ 
with $\ho_i>0$, placing the value of $\ho_i$ in a small circle at the point 
$(i,\hh_i)$. 

\subsubsection{Example}
An example is given in Figure {\bf \ref{tTypicalA11a}}.

\begin{figure}[ht]
\begin{center}
\psset{yunit=0.32cm,xunit=0.28cm}
\begin{pspicture}(-2,-1)(32,6.5)
\rput[bl](0,0){
\psset{linewidth=0.25pt,linestyle=dashed, dash=2.5pt 1.5pt,linecolor=gray}
\multips(0,1)(0,1){5}{\psline{-}(0,0)(32,0)}
\multips(0,0)(0,0){5}{\psline{-}(0,0)(32,0)}
\multips(1,0)(1,0){31}{\psline{-}(0,0)(0,5)}
\multips(0,0)(0,0){31}{\psline{-}(0,0)(0,5)}
\rput[r](-0.5,5){\scriptsize $5$}
\rput[r](-0.5,4){\scriptsize $4$}
\rput[r](-0.5,3){\scriptsize $3$}
\rput[r](-0.5,2){\scriptsize $2$}
\rput[r](-0.5,1){\scriptsize $1$}
\rput[r](-0.5,0){\scriptsize $0$}
\psset{linewidth=0.7pt,fillstyle=none,linestyle=solid,linecolor=black}
\psline(0,3)(2,1)(4,3)(7,0)(8,0)(9,1)(10,0)(15,5)(17,3)(18,4)(22,0)(24,0)
       (26,2)(28,0)(32,0)
\psset{dotsize=3mm,fillstyle=none,fillcolor=white,dotstyle=o}
\rput(3,2){\psdot(0,0)\rput(0,0){\scriptsize $3$}}
\rput(6,1){\psdot(0,0)\rput(0,0){\scriptsize $1$}}
\rput(7,0){\psdot(0,0)\rput(0,0){\scriptsize $2$}}
\rput(18,4){\psdot(0,0)\rput(0,0){\scriptsize $3$}}
\rput(20,2){\psdot(0,0)\rput(0,0){\scriptsize $1$}}
\rput(29,0){\psdot(0,0)\rput(0,0){\scriptsize $4$}}
}
\end{pspicture}
\end{center}
\caption{\it 
Typical $\thbress\in\tAdset{5}{3}$
}
\label{tTypicalA11a}
\end{figure}

The number of peaks $\npt(\thbress)$ and weight $\wt(\thbress)$ of
$\thbress=(\hbress,\ho)\in\Adb$ are defined by

\begin{subequations}\label{Eq:tdefs}
\begin{align}
\label{Eq:thtdef}
\npt(\thbress)&=\npt(\hbress)+\ell(\ho),\\
\label{Eq:twtdef}
\wt(\thbress)&=\wt(\hbress)+|\ho|,
\end{align}
\end{subequations}

\noindent where $\ell(\ho)=\sum_{i=0}^\infty\ho_i$ and $|\ho|=\sum_{i=1}^\infty i\ho_i$.
Note that $\ho$ is, in effect, a partition, with $\ho_i$ giving the multiplicity of the 
part $i$ for $i>0$, and $|\ho|$ giving its weight. However, $\ell(\ho)$ is a variant on 
the standard definition of partition length because, here, zero parts are counted.
We then define the set
$\tAdb(a)=\{\thbress\in\tAdb\,|\,\npt(\thbress)=a\}$
and the generating functions

\begin{equation}
\label{Eq:tAdbGF}
\tAdbGF
= \sum_{\thbress\in\tAdb} q^{\wt(\thbress)},\qquad
\tAdbGF(a)
=\sum_{\thbress\in\tAdb(a)}
q^{\wt(\thbress)}
\end{equation}

\begin{lemma}\label{A2}
For $0\le b\le k$,
\begin{equation}\label{Eq:A2}
\tAdbGF(a)=\sum_{n=0}^a \frac{1}{(q;q)_{a-n}} \AdbGF(n)
\end{equation}
\end{lemma}

\Proof The required expression is the generating function for decorated $\B$-paths
$\thbress=(\hh,\ho)$ with fixed number of peaks $\npt(\thbress)=a$. By \eqref{Eq:thtdef},
$0\le\npt(\hh)\le a$ with $\ell(\ho)=a-\npt(\hh)$. Therefore,

\begin{equation}
\tAdbGF(a)=\sum_{n=0}^a
\AdbGF(n)\sum_{\ho|\ell(\ho)=a-n} q^{|\ho|}
\end{equation}

\noindent The second sum here is given by $(q;q)_{a-n}^{-1}$, this being the generating 
function for partitions with at most $(a-n)$ parts. Then, \eqref{Eq:A2} immediately follows.
\cqfd

\subsection{The $\B$-transform}
\label{A1Tran}

Here, we describe a way to transform a decorated $\B$-path $\thbress\in\tAdset{k}{b}$ 
to obtain a (non-decorated) $\B$-path $\hbress'\in\Adset{k+1}{b}$. This action extends 
the notion of {\it \lq volcanic uplift\rq\,} described in \cite{bressoud.1989}, and we 
refer to it as a $\B$-transform.

Let $\thbress=(\hbress,\ho)$, and define $\hp=(\hp_0,\hp_1,\hp_2, \cdots)$ by setting 
$\hp_i=\ho_i+1$ if $i$ is a peak, and $\hp_i=\ho_i$ if $i$ is not a peak. Note that 
$\sum_{i=0}^\infty \hp_i=\npt(\thbress)$. The $\B$-path $\hbress'$ is obtained from 
$\hbress$ simply by, for $i= \cdots,2,1,0$, inserting $\hp_i$ NE-SE pairs at position 
$(i,\hh_i)$. In effect, each peak of $\hbress$ gives rise to a peak of $\hbress'$ of 
height one greater. By regarding $\ho_i>0$ deaks as a sequence of $\ho_i$ degenerate 
peaks at $(i,\hh_i)$, this statement applies to them as well.

To illustrate the action of the $\B$-transform, it maps the decorated $\B$-path $\thbress$ 
of Figure {\bf \ref{BBefore}} to the $\B$-path $\hbress'$ of Figure {\bf \ref{BAfter}}.
Note that in this example $\npt(\hbress')=\npt(\thbress)=10$ and 
$\wt(\thbress)-\wt(\hbress')=100$.

\begin{figure}[ht]
\begin{center}
\psset{yunit=0.32cm,xunit=0.28cm}
\begin{pspicture}(-2,-1)(21,4.5)
\rput[bl](0,0){
\psset{linewidth=0.25pt,linestyle=dashed, dash=2.5pt 1.5pt,linecolor=gray}
\multips(0,1)(0,1){3}{\psline{-}(0,0)(21,0)}
\multips(0,0)(0,0){3}{\psline{-}(0,0)(21,0)}
\multips(1,0)(1,0){20}{\psline{-}(0,0)(0,3)}
\multips(0,0)(0,0){20}{\psline{-}(0,0)(0,3)}
\rput[r](-0.8,3){\scriptsize $3$}
\rput[r](-0.8,2){\scriptsize $2$}
\rput[r](-0.8,1){\scriptsize $1$}
\rput[r](-0.8,0){\scriptsize $0$}
\psset{linewidth=0.7pt,fillstyle=none,linestyle=solid,linecolor=black}
\psline(0,2)(1,1)(3,3)(6,0)(7,0)(8,1)(9,0)(10,0)(12,2)(13,1)(15,3)(18,0)(21,0)
\psset{dotsize=3mm,fillstyle=none,fillcolor=white,dotstyle=o}
\rput(0,2){\psdot(0,0)\rput(0,0){\scriptsize $2$}}
\rput(4,2){\psdot(0,0)\rput(0,0){\scriptsize $1$}}
\rput(9,0){\psdot(0,0)\rput(0,0){\scriptsize $1$}}
\rput(15,3){\psdot(0,0)\rput(0,0){\scriptsize $2$}}
}
\end{pspicture}
\end{center}
\caption{\it 
Before $\thbress\in\tAdset{3}{2}(10)$
}
\label{BBefore}
\end{figure}

\begin{figure}[ht]
\begin{center}
\psset{yunit=0.32cm,xunit=0.28cm}
\begin{pspicture}(-2,-1)(41,5.5)
\rput[bl](0,0){
\psset{linewidth=0.25pt,linestyle=dashed, dash=2.5pt 1.5pt,linecolor=gray}
\multips(0,1)(0,1){4}{\psline{-}(0,0)(41,0)}
\multips(0,0)(0,0){4}{\psline{-}(0,0)(41,0)}
\multips(1,0)(1,0){40}{\psline{-}(0,0)(0,4)}
\multips(0,0)(0,0){40}{\psline{-}(0,0)(0,4)}
\rput[r](-0.8,4){\scriptsize $4$}
\rput[r](-0.8,3){\scriptsize $3$}
\rput[r](-0.8,2){\scriptsize $2$}
\rput[r](-0.8,1){\scriptsize $1$}
\rput[r](-0.8,0){\scriptsize $0$}
\psset{linewidth=0.7pt,fillstyle=none,linestyle=solid,linecolor=black}
\psline(0,2)(1,3)(2,2)(3,3)(5,1)(8,4)(10,2)(11,3)(14,0)(15,0)(17,2)
       (19,0)(20,1)(21,0)(22,0)(25,3)(27,1)(30,4)(31,3)(32,4)(33,3)
       (34,4)(38,0)(41,0)
}
\end{pspicture}
\end{center}
\caption{\it 
Transformed $\hbress'\in\Adset{4}{2}(10)$
}
\label{BAfter}
\end{figure}

\begin{lemma}\label{A1param}
For $0\le b\le k$, let $\hbress'\in\Adset{k+1}{b}$ be obtained from the action 
of a $\B$-transform on $\thbress\in\tAdset{k}{b}(a)$.  Then

\begin{subequations}\label{Eq:A1param}
\begin{align}
\label{Eq:A1parama}
\npt(\hbress')&=a,\\
\label{Eq:A1paramb}
\wt(\hbress')&=\wt(\thbress)+a^2
\end{align}
\end{subequations}
\end{lemma}

\Proof
Altogether, $\thbress$ has $a$ peaks and deaks. The first expression holds because
each of these gives rise to a (genuine) peak of $\hbress'$, with all peaks of 
$\hbress'$ obtained thus. Let $p_1,p_2, \cdots,p_a$ be, in non-decreasing order, the 
peaks and deaks of $\thbress$. The peaks of $\hbress'$ are then at positions
$p_1+1,p_2+3, \cdots,p_a+2a-1$. From \eqref{Eq:BressWtDef}, it follows that

\begin{equation}
\wt(\hbress')=\sum_{i=1}^a p_i+1+3+5+ \cdots+(2a-1),
\end{equation}

\noindent which immediately gives \eqref{Eq:A1paramb}.
\cqfd

\begin{lemma}\label{A1}
For $0\le b\le k$,
\begin{equation}\label{Eq:A1}
\AGF{k+1}{b}(a)=q^{a^2}\,\tAdbGF(a)
\end{equation}
\end{lemma}

\Proof
This immediately follows from Lemma {\bf \ref{A1param}} once it is shown that the 
$\B$-transform provides a bijection between $\tAdset{k}{b}(a)$ and $\Adset{k+1}{b}(a)$.
This is so because if $\hbress\in\Adset{k+1}{b}(a)$, the unique $\thbress\in\tAdset{k}{b}(a)$ 
from which it arose is obtained by removing the two edges next to each peak, and taking account 
of the multiplicities of the downgraded peaks to give peaks and deaks.
\cqfd

\begin{lemma}\label{A1GF}
If $0\le b<k$ then
\begin{subequations}\label{Eq:A1GF}
\begin{align}
\label{Eq:A1GFa}
\tAGF{k}{b}(a)&=\sum_{n=0}^a \frac{q^{n^2}}{(q;q)_{a-n}}\,\tAGF{k-1}{b}(n)\\
\intertext{If $k>0$ then}
\label{Eq:A1GFb}
\tAGF{k}{k}(a)&=\sum_{n=0}^a \frac{q^{n(n+1)}}{(q;q)_{a-n}}\,\tAGF{k-1}{k-1}(n)
\end{align}
\end{subequations}
\end{lemma}

\Proof
The first expression here results from substituting the $k\to k-1$ case of 
\eqref{Eq:A1} into \eqref{Eq:A2}. The second expression results from first 
substituting \eqref{Eq:U1} into the $b=k$ case of \eqref{Eq:A2}, and then 
applying the $k\to k-1$ and $b\to k-1$ case of \eqref{Eq:A1}.
\cqfd

\section{Proof of the Andrews-Gordon identities}
\label{andrews.gordon}
\noindent {\it
We prove the Andrews-Gordon identities by using abaci to map cylindric partitions 
to $\B$-paths, then use $\B$-transforms.
}

\subsection{The sum side of the Andrews-Gordon identities}
\label{Chain}

We now obtain the desired fermionic-type expression for $\tAdbGF(a)$ by 
concatenating together a sequence of $B$-transforms.

\begin{theorem}\label{Achain}
If $k\ge1$ and $0\le b\le k$ then

\begin{equation}\label{Eq:Achain}
\tAGF{k}{b}(a)=
\sum_{a\ge n_k\ge n_{k-1}\ge \cdots\ge n_1\ge0}
\frac{q^{n_1^2+n_2^2+ \cdots+n_k^2
      +n_{1}+n_{2}+ \cdots+n_{b}}}
     {(q;q)_{a-n_k}(q;q)_{n_k-n_{k-1}}(q;q)_{n_{k-1}-n_{k-2}}
      \cdots(q;q)_{n_{2}-n_1}(q;q)_{n_1}}
\end{equation}
\end{theorem}

\Proof
Each $(\hbress,\ho)\in\tAdset{0}{0}(a)$ has $\hbress=(0,0,0, \cdots)$.
Because $\ho$ here is a partition unconstrained apart from having
no more than $a$ parts, $\tAGF{0}{0}(a)=(q;q)_a^{-1}$. The two cases 
of \eqref{Eq:A1GF} then imply that

\begin{subequations}
\begin{align}
\label{Eq:AchainPf1a}
\tAGF{1}{0}(a)&=\sum_{n_1=0}^a\frac{q^{n_1^2}}{(q;q)_{a-n_1}(q;q)_{n_1}}\,,\\
\label{Eq:AchainPf1b}
\tAGF{1}{1}(a)&=\sum_{n_1=0}^a\frac{q^{n_1^2+n_1}}{(q;q)_{a-n_1}(q;q)_{n_1}}
\end{align}
\end{subequations}

\noindent These are the $k=1$ cases of \eqref{Eq:Achain}. We now proceed to 
prove \eqref{Eq:Achain} by induction on $k$, by assuming that \eqref{Eq:Achain} 
holds for $k\to k-1$ and $0\le b\le k-1$. We consider the case $0\le b\le k-1$ 
and the case $b=k$ separately.

\subsubsection{The case $b\le k-1$} Using \eqref{Eq:A1GFa}, and the induction hypothesis 
yields

\begin{equation}
\begin{split}
\tAGF{k}{b}(a)
&= \sum_{n_k=0}^a \frac{q^{n_k^2}}{(q;q)_{a-n_k}}\, \tAGF{k-1}{b}(n_k)\\
&= \sum_{n_k=0}^a \frac{q^{n_k^2}}{(q;q)_{a-n_k}}
\sum_{n_k\ge n_{k-1}\ge \cdots\ge n_1\ge0}
\frac{q^{n_1^2+n_2^2+ \cdots+n_{k-1}^2
      +n_{1}+n_{2}+ \cdots+n_{b}}}
     {(q;q)_{n_k-n_{k-1}}(q;q)_{n_{k-1}-n_{k-2}}
      \cdots(q;q)_{n_{2}-n_1}(q;q)_{n_1}},
\end{split}
\end{equation}

\noindent which is the $b\le k-1$ case of \eqref{Eq:Achain}, as desired.

\subsubsection{The case $b=k$} Using \eqref{Eq:A1GFb}, and the induction hypothesis 
yields

\begin{equation}
\begin{split}
\tAGF{k}{k}(a)
&= \sum_{n_k=0}^a \frac{q^{n_k^2+n_k}}{(q;q)_{a-n_k}}\, \tAGF{k-1}{k-1}(n_k)\\
&= \sum_{n_k=0}^a \frac{q^{n_k^2+n_k}}{(q;q)_{a-n_k}}
\sum_{n_k\ge n_{k-1}\ge \cdots\ge n_1\ge0}
\frac{q^{n_1^2+n_2^2+ \cdots+n_{k-1}^2
      +n_{1}+n_{2}+ \cdots+n_{k-1}}}
     {(q;q)_{n_k-n_{k-1}}(q;q)_{n_{k-1}-n_{k-2}}
      \cdots(q;q)_{n_{2}-n_1}(q;q)_{n_1}},
\end{split}
\end{equation}

\noindent which is the $b=k$ case of \eqref{Eq:Achain}, as desired.
\cqfd

\subsection{Characterising decorated $\B$-paths}
\label{CharPath}

For $\thbress\in(\hbress,\ho)\in\tAdb(a)$, let $\{i_j\}_{j=1}^a$ be the set of 
peaks and deaks of $\thbress$ with $0\le i_a\le i_{a-1}\le \cdots\le i_1$
(for each value that is repeated, all but one of them corresponds to a deak),
and let $b_j=h_{i_j}$ for $1\le j\le a$. In addition, for convenience, set 
$i_{a+1}=0$, $b_{a+1}=b$, and $\delta_j=i_j-i_{j+1}$ for $1\le j\le a$.
Then, \eqref{Eq:twtdef} gives

\begin{equation}
\label{Eq:CharPath}
\wt(\thbress) =\sum_{j=1}^a i_j =\sum_{j=1}^a j\delta_j
\end{equation}

We claim that $\thbress$ is completely determined by the points $\{(i_j,b_j)\}_{j=1}^a$.
This follows because, for $1\le j\le a$, there is no peak strictly between $i_{j+1}$ and 
$i_j$, and therefore the sequence of $\delta_j$ edges of the path $\hbress$ can only take 
one form, namely, a sequence of SE edges followed by a sequence of E edges at height 0 
followed by 
a sequence of NE edges, with any of these sequences of zero length. Thus $\thbress$ is 
determined uniquely. Moreover, depending on whether there are non-zero or zero E edges, 
either

\begin{subequations}\label{Eq:DeltaCons}
\begin{align}
&\delta_j> b_j+b_{j+1}\text{ or}\\
&|b_j-b_{j+1}|\le\delta_j\le b_j+b_{j+1}
\text{ with }
\delta_j-|b_j-b_{j+1}|\in2\ZZ
\end{align}
\end{subequations}

\noindent These conditions are sufficient to determine whether $\{(i_j,b_j)\}_{j=1}^a$ 
corresponds to a $\thbress\in\tAdb(a)$.

\subsection{Bijection from cylindric partitions to decorated $\B$-paths}
\label{BijDBpaths}

For $d$ odd, set $d=2k+1$.
Then, for $\Lambda=[d-x,x]\in\wtlatd{2}{d}$,
let $b=\max\{k-x,x-k-1\}$ so that $0\le b\le k$.
In what follows, for $\pi\in\CPX{2}{\Lambda}(a)$, we use
$\boldLambda(\pi)=(\cdots,\Lambda^{(3)},\Lambda^{(2)},\Lambda^{(1)})$
and $\bolddelta(\pi)=(\cdots,\delta_3,\delta_2,\delta_1)$
to construct an element $\thbress\in\tAdset{k}{b}(a)$
by specifying the set of points $\{(i_j,b_j)\}_{j=1}^a$
of its $a$ peaks and deaks.
We then show that this defines a weight-preserving bijection between
$\CPX{2}{\Lambda}(a)$ and $\tAdset{k}{b}(a)$.

First set $i_j=\sum_{m=j}^a \delta_m$ for $1\le j\le a$.
Then, for $j>0$, let $x_j$ be such that $\Lambda^{(j)}=[d-x_j,x_j]$,
and obtain $b_j$ by
\begin{equation}\label{Eq:bjDef}
b_j=
\begin{cases}
k-x_j&\text{if $x_j\le k$,}\\
x_j-k-1&\text{if $x_j>k$,}
\end{cases}
\end{equation}

\noindent {\it i.e.} $b_j=\max\{k-x_j,x-k_j-1\}$.
Note that $0\le b_j\le k$ with $b_{a+1}=b$.
Making use of \eqref{Eq:Distsr=2a}, we then find that if both
$x_{j+1}\le k$ and $x_j\le k$, or both
$x_{j+1}>k$ and $x_j>k$, then
\begin{subequations}\label{Eq:deltaConds}
\begin{align}
\label{Eq:deltaConds1}
\delta_j&\in 2\ZZp+|b_{j+1}-b_j|.\\
\intertext{Otherwise,}
\label{Eq:deltaConds2}
\delta_j&\in 2\ZZp+b_{j+1}+b_j+1
\end{align}
\end{subequations}

\noindent After noting that one of $|b_{j+1}-b_j|$ and $(b_{j+1}+b_j+1)$ 
is even and one odd, we see that $\{(i_j,b_j)\}_{j=1}^a$ satisfies the 
conditions \eqref{Eq:DeltaCons}, thereby determining a unique element 
$\thbress\in\tAdset{k}{b}(a)$.

That the map so defined is a bijection follows because,
given values $\{(i_j,b_j)\}_{j=1}^a$ that satisfy the conditions in
Section {\bf \ref{CharPath}}, values $x_a,x_{a-1}, \cdots,x_1$
with $0\le x_j\le 2k+1$ are uniquely determined by reversing the
above construction.
This is done by noting that with $i_{a+1}=0$, $b_{a+1}=b$, and
$\delta_j=i_j-i_{j+1}$ for $1\le j\le a$,
\eqref{Eq:deltaConds} implies that
\begin{equation}\label{Eq:Logic}
(x_{j+1}\le k \Leftrightarrow x_j\le k)
\quad\Leftrightarrow\quad
\delta_j\equiv b_{j+1}+b_j\pmod2,
\end{equation}
for $1\le j\le a$.
Then, with $x_{a+1}$ such that $\Lambda=[d-x_{a+1},x_{a+1}]$,
each value $x_j$ is determined from $b_j$, $b_{j+1}$, $\delta_j$ and
$x_{j+1}$ using first \eqref{Eq:Logic} to determine whether
$x_j\le k$ or $x_j>k$, and then making use of the appropriate
case of \eqref{Eq:bjDef}.
On setting $\Lambda^{(j)}=[d-x_j,x_j]$ for $1\le j\le a$,
and $\Lambda^{(j)}=\Lambda$ and $\delta_j=0$ for $j>a$,
$\pi\in\CPX{2}{\Lambda}(a)$ is uniquely determined such that
$\boldLambda(\pi)
  =(\cdots,\Lambda^{(3)},\Lambda^{(2)},\Lambda^{(1)})$
and $\bolddelta(\pi)=(\cdots,\delta_3,\delta_2,\delta_1)$,
thereby demonstrating the map to be bijective.

From \eqref{Eq:CharPath} and \eqref{Eq:piWtDef},
$\wt(\thbress)=\sum_{j=1}^a j\delta_j=|\pi|$, thereby showing
the bijection to be weight-preserving.

\subsection{The product side of the Andrews-Gordon identities}
The bijection of the previous subsection implies the following result,

\begin{lemma}\label{WtPresOdd}
For $k\ge0$, let $\Lambda\in\wtlatd{2}{2k+1}$
be of the form $\Lambda=[k+1+b,k-b]$ or $\Lambda=[k-b,k+1+b]$
for $0\le b\le k$.
Then, for each $a\ge0$,
\begin{equation}
\CPgfX{2}{\Lambda}(a)=\tAGF{k}{b}(a).
\end{equation}
\end{lemma}

\noindent
Using this result, and setting $b=k-i$ and $n_j=\lambda_{k+1-j}$, 
the odd $d$ case of Theorem {\bf \ref{CPAG}} follows from Theorem 
{\bf \ref{Achain}}.

\section{Translating the proof in terms of abaci}
\label{translation}

\noindent {\it 
We consider how the proof of Theorem {\bf \ref{Achain}} translates 
to the abaci through the bijection of section {\bf \ref{BijDBpaths}}.
}

\subsection{The $\B$-transform revisited}
Expression \eqref{Eq:Achain} results from the application of $(k+1)$ $\B$-transforms.
At the $i$-th stage ($0\le i\le k$), the factor $(q;q)_{n_{i+1}-n_{i}}^{-1}$ in
\eqref{Eq:Achain} corresponds to the placement of degenerate peaks on a $\B$-path $\hbress$.
When translated to abaci, these correspond to the placement of yokes of varying shapes.

\subsection{Interpretation of proof in terms of yoke moves}
\label{YokeMoves}

The generating function for $\tAGF{k}{b}(a)$ given by \eqref{Eq:Achain}
is constructed by alternating a sequence of $\B$-transforms
(encoded in \eqref{Eq:A1})
with multiplication by $(q;q)^{-1}_{n_{i+1}-n_i}$ for
fixed $n_{i+1}$ and summed over $n_i$ with $0\le n_i\le n_{i+1}$
(encoded in \eqref{Eq:A2}).
Here $1\le i\le k$ and we take $n_{k+1}=a$.
It is instructive to interpret these transformations between generating
functions for decorated $\B$-paths in terms of
the abaci of the cylindric partitions, through the above bijection.

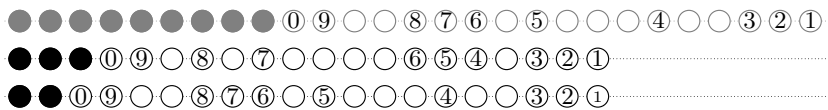
\begin{figure}[ht]
\begin{center}
\psset{yunit=5.0mm,xunit=4.0mm}
\begin{pspicture}(-8,-0.5)(20,3.0)
\psset{linewidth=0.25pt,linestyle=dotted,dotsep=1.0pt,linecolor=black}
\multips(0,0)(0,1){2}{\psline{-}(-7.5,0)(19.5,0)}
\psset{linewidth=0.25pt,linestyle=dotted,dotsep=1.0pt,linecolor=gray}
\multips(0,2)(0,1){1}{\psline{-}(-7.5,0)(19.5,0)}
\psset{dotsize=3mm,fillstyle=solid,fillcolor=black,dotstyle=o}
\psset{linecolor=black}
\multips(-7,0)(1,0){2}{\psdots(0,0)}
\multips(-7,1)(1,0){3}{\psdots(0,0)}
\psset{linecolor=gray,fillcolor=gray}
\multips(-7,2)(1,0){9}{\psdots(0,0)}
\psset{dotsize=3mm,fillstyle=solid,fillcolor=white,dotstyle=o}
\psset{linecolor=black}
\multips(-5,0)(1,0){18}{\psdots(0,0)}
\multips(-4,1)(1,0){17}{\psdots(0,0)}
\psset{linecolor=gray}
\multips(2,2)(1,0){18}{\psdots(0,0)}
\rput(12,0){$\scriptscriptstyle 1$}\rput(12,1){\scriptsize $1$}
  \rput(19,2){\scriptsize $1$}
\rput(11,0){\scriptsize $2$}\rput(11,1){\scriptsize $2$}
  \rput(18,2){\scriptsize $2$}
\rput(10,0){\scriptsize $3$}\rput(10,1){\scriptsize $3$}
  \rput(17,2){\scriptsize $3$}
\rput(7,0){\scriptsize $4$}\rput(8,1){\scriptsize $4$}
  \rput(14,2){\scriptsize $4$}
\rput(3,0){\scriptsize $5$}\rput(7,1){\scriptsize $5$}
  \rput(10,2){\scriptsize $5$}
\rput(1,0){\scriptsize $6$}\rput(6,1){\scriptsize $6$}
  \rput(8,2){\scriptsize $6$}
\rput(0,0){\scriptsize $7$}\rput(1,1){\scriptsize $7$}
  \rput(7,2){\scriptsize $7$}
\rput(-1,0){\scriptsize $8$}\rput(-1,1){\scriptsize $8$}
  \rput(6,2){\scriptsize $8$}
\rput(-4,0){\scriptsize $9$}\rput(-3,1){\scriptsize $9$}
  \rput(3,2){\scriptsize $9$}
\rput(-5,0){\scriptsize $0$}\rput(-4,1){\scriptsize $0$}
  \rput(2,2){\scriptsize $0$}
\end{pspicture}
\end{center}
\caption{\it 
$r=2$, $d=7$ abacus (for $\Lambda=6\Lambda_0+\Lambda_1$)
}
\label{TypicalAbacusr=2b}
\end{figure}

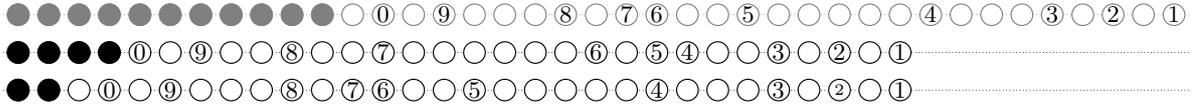
\begin{figure}[ht]
\begin{center}
\psset{yunit=5.0mm,xunit=4.0mm}
\begin{pspicture}(-8,-0.5)(32,3.0)
\psset{linewidth=0.25pt,linestyle=dotted,dotsep=1.0pt,linecolor=black}
\multips(0,0)(0,1){2}{\psline{-}(-7.5,0)(31.5,0)}
\psset{linewidth=0.25pt,linestyle=dotted,dotsep=1.0pt,linecolor=gray}
\multips(0,2)(0,1){1}{\psline{-}(-7.5,0)(31.5,0)}
\psset{dotsize=3mm,fillstyle=solid,fillcolor=black,dotstyle=o}
\psset{linecolor=black}
\multips(-7,0)(1,0){2}{\psdots(0,0)}
\multips(-7,1)(1,0){4}{\psdots(0,0)}
\psset{linecolor=gray,fillcolor=gray}
\multips(-7,2)(1,0){11}{\psdots(0,0)}
\psset{dotsize=3mm,fillstyle=solid,fillcolor=white,dotstyle=o}
\psset{linecolor=black}
\multips(-5,0)(1,0){28}{\psdots(0,0)}
\multips(-3,1)(1,0){26}{\psdots(0,0)}
\psset{linecolor=gray}
\multips(4,2)(1,0){28}{\psdots(0,0)}
\rput(22,0){\scriptsize $1$}\rput(22,1){\scriptsize $1$}
  \rput(31,2){\scriptsize $1$}
\rput(20,0){$\scriptscriptstyle 2$}\rput(20,1){\scriptsize $2$}
  \rput(29,2){\scriptsize $2$}
\rput(18,0){\scriptsize $3$}\rput(18,1){\scriptsize $3$}
  \rput(27,2){\scriptsize $3$}
\rput(14,0){\scriptsize $4$}\rput(15,1){\scriptsize $4$}
  \rput(23,2){\scriptsize $4$}
\rput(8,0){\scriptsize $5$}\rput(14,1){\scriptsize $5$}
  \rput(17,2){\scriptsize $5$}
\rput(5,0){\scriptsize $6$}\rput(12,1){\scriptsize $6$}
  \rput(14,2){\scriptsize $6$}
\rput(4,0){\scriptsize $7$}\rput(5,1){\scriptsize $7$}
  \rput(13,2){\scriptsize $7$}
\rput(2,0){\scriptsize $8$}\rput(2,1){\scriptsize $8$}
  \rput(11,2){\scriptsize $8$}
\rput(-2,0){\scriptsize $9$}\rput(-1,1){\scriptsize $9$}
  \rput(7,2){\scriptsize $9$}
\rput(-4,0){\scriptsize $0$}\rput(-3,1){\scriptsize $0$}
  \rput(5,2){\scriptsize $0$}
\end{pspicture}
\end{center}
\caption{\it 
$r=2$, $d=9$ abacus (for $\Lambda=7\Lambda_0+2\Lambda_1$)
}
\label{TypicalAbacusr=2c}
\end{figure}

In the first place, \eqref{Eq:A1} with $k\to i-1$ and $a\to n_i$ is
interpreted in terms of each yoke changing shape:
for the rightmost $n_{i}$ yokes, a $[m_0,m_1]$-yoke
changes to either a $[m_0,m_1+2]$-yoke or a $[m_0,m_1+2]$-yoke
depending on whether $m_0<m_1$ or $m_0\ge m_1$ respectively;
other than those rightmost $n_{i}$, the Dirac sea of $[i+b,i-1-b]$-yokes
become $[i+1+b,i-b]$-yokes.
In addition, the distance between the $j+1$th and $j$th yokes
increases by 2 for $1\le j<n_i$, increases by 1 for $j=n_i$,
and remains 0 for $j>n_i$.
This process is demonstrated by the transition between the
abacus in Figure {\bf \ref{TypicalAbacusr=2b}}
and that in Figure {\bf \ref{TypicalAbacusr=2c}}.
In these abaci, we have labelled the beads in the $j$th yoke with $j$,
where $1\le j\le n_i=10$ (substituting the label `10' by `0').
Note that the abaci in Figures {\bf \ref{TypicalAbacusr=2b}}
and {\bf \ref{TypicalAbacusr=2c}} correspond respectively to the
decorated path in Figure {\bf \ref{BBefore}} and the
undecorated path in Figure {\bf \ref{BAfter}}.

To interpret \eqref{Eq:A2}, consider an element $\hbress\in\AGF{i}{b}(n_i)$
and its corresponding abacus $\pi$.
All but the rightmost $n_i$ yokes of $\pi$ belong to the Dirac sea
and have weight $[i+1+b,i-b]$.
The rightmost $n_i$ yokes are unconstrained, apart from each being
of a weight of level $2i+1$, and being separated from one-another
by at least two vacancies.

As described in Section {\bf \ref{DBpaths}}, multiplication by 
$(q;q)^{-1}_{n_{i+1}-n_i}$ enumerates all sets of positions of $n_{i+1}-n_i$ 
deaks along the path $\hbress$, the positions being encoded in a sequence 
$\ho=(\ho_0,\ho_1,\ho_2, \cdots)$ for which $\ell(\ho)=n_{i+1}-n_i$.
In terms of abaci, this corresponds to enumerating all possible abaci 
$\pi'$ obtained from $\pi$ by inserting $n_{i+1}-n_i$ yokes between the 
rightmost $n_i+1$ yokes of $\pi$.
For a particular $\ho=(\ho_0,\ho_1,\ho_2, \cdots)$, for each
$j\ge0$ for which $\ho_j>0$, the $\ho_j$ yokes are inserted
such that there are exactly $j$ vacancies to their left.
Their weight depends on the value of $\hh_j$:
either $[i+1+\hh_j,i-\hh_j]$ or $[i-\hh_j,i+1+\hh_j]$,
with the choice between these two determined by \eqref{Eq:rDist}.

Alternatively, all the abaci $\pi'$ may be generated by starting
with $\pi$ and performing a sequence of `moves' on the yokes
numbered $n_{i}+1$ to $n_{i+1}$.
Note that $\pi$ itself is the abacus corresponding to
$\ho=(n_{i+1}-n_{i},0,0,0, \cdots)$.
Each move corresponds to changing the deak positions from one
$\ho=(\ho_0, \cdots,\ho_j,\ho_{j+1}, \cdots)$
to $(\ho_0, \cdots,\ho_j-1,\ho_{j+1}+1, \cdots)$ for some $j$.
Such a move on the abacus is simply the shifting of one bead
from the yoke corresponding to the deak at $j$ one position to the right.
Which of the two beads moved may be determined from $\hbress$.
In brief, if only one of the two beads can be moved, it does so.
If both beads can move then the one chosen is such that the
resulting yoke weight $[m'_0,m'_1]$ has $|m'_0-m'_1|$ a minimum.
If neither bead can move, then the yoke is necessarily
immediately adjacent to another of the same weight.
The move (and subsequent moves) are then performed on the beads
of this yoke instead.

\section{Proof of the Bressoud identities}
\label{bressoud}

\noindent {\it
We prove Bressoud's identities using a direct extension of the proof of 
the Andrews-Gordon identities used in section {\bf \ref{andrews.gordon}}.
}

\subsection{Even $\B$-paths}
\label{EBpaths}

For $0\le b\le k$ define $\Bdb$ to be the subset of $\Adb$
comprising all $\hbress=(h_0,h_1,h_2, \cdots)$ for which
if $h_i=k$ then $i\equiv k-b\pmod2$.
These paths were also originally considered in \cite{bressoud.1989}.
It is easy to see that this condition implies that there are an
even number of horizontal segments to the left of each peak $i$
for which $h_i=k$.
Note that the path $\hbress\in\Adset53$ of Figure {\bf \ref{TypicalA11a}}
is not an element of $\Bdset53$.
We then define the set
$\Bdb(a)=\{\hbress\in\Bdb\,|\,\npt(\hbress)=a\}$
and the generating functions
\begin{equation}\label{Eq:BdbGF}
\BdbGF
= \sum_{\hbress\in\Bdb} q^{\wt(\hbress)},\qquad
\BdbGF(a)
= \sum_{\hbress\in\Bdb(a)} q^{\wt(\hbress)}
\end{equation}

We have the following analogue of Lemma {\bf \ref{U1}},

\begin{lemma}\label{U2}
For $k>0$,
\begin{equation}\label{Eq:U2}
\BGF{k}{k}(a)=q^a\,\BGF{k}{k-1}(a)
\end{equation}
\end{lemma}

\Proof Each $\hbress\in\Bdset{k}{k}(a)$ is necessarily such that
$\hh_0=k$ and $\hh_1=k-1$. Therefore $i=1$ not a peak.
Removing the first segment of $\hbress$ therefore results in
an element $\hbress'\in\Adset{k}{k-1}(a)$ in which each peak is one
less than the corresponding peak in $\hbress$.
Thus $\npt(\hbress')=\npt(\hbress)=a$
and $\wt(\hbress')=\wt(\hbress)-a$.
Moreover, because the parity of the startpoint has changed,
$\hbress'\in\Bdset{k}{k-1}(a)$.
Then, because every element $\hbress'\in\Bdset{k}{k-1}(a)$ arises
from a unique $\hbress\in\Bdset{k}{k}(a)$ in this way,
\eqref{Eq:U2} immediately follows.
\cqfd

The paths $\hbress\in\Bdb$ are now decorated much as before.
For $k>0$, we define $\tBdb$ to be the subset of $\tAdb$
comprising all decorated $\B$-paths
$\thbress=(\hbress,\ho)$ for which $\hbress\in\Bdb$
and $\ho$ is any sequence $\ho=(\ho_0,\ho_1,\ho_2, \cdots)$
of non-negative integers,
for which there exists $M\ge0$ such that $\ho_i=0$ for $i>M$.
For the $k=0$ case (and only this case), we restrict the $\ho$ 
to sequences
$\ho=(\ho_0,\ho_1,\ho_2, \cdots)$ for which $\ho_i=0$ for $i$ odd
\footnote{\, 
This means that for all $k\ge0$, $\tBdb$ comprises
all decorated paths $\thbress$ whose peaks and deaks $i$
for which $h_i=k$ satisfy $i\equiv k-b\pmod2$.
}.

We then define the set
$\tBdb(a)=\{\thbress\in\tBdb\,|\,\npt(\thbress)=a\}$
and the generating functions
\begin{equation}\label{Eq:tBdbGF}
\tBdbGF
= \sum_{\thbress\in\tBdb} q^{\wt(\thbress)},\qquad
\tBdbGF(a)
=\sum_{\thbress\in\tBdb(a)}
q^{\wt(\thbress)}.
\end{equation}

\begin{lemma}\label{B2}
For $k>0$ and $0\le b\le k$,
\begin{equation}\label{Eq:B2}
\tBdbGF(a)=\sum_{n=0}^a \frac{1}{(q;q)_{a-n}} \BdbGF(n)
\end{equation}
\end{lemma}

\Proof
This is proved in exactly the same way as Lemma {\bf \ref{A2}}.
\cqfd

\subsection{$\B$-transform for even $\B$-paths}
\label{B1Tran}

For $\thbress\in\tAdset{k}{b}$, the action of the $\B$-transform of Section 
{\bf \ref{A1Tran}} yields $\hbress'\in\Adset{k+1}{b}$ each of whose peaks is
of the opposite parity to that of the corresponding peak of $\thbress$.
In particular, the $\B$-transform defines a map from
$\thbress\in\tBdset{k}{b}$ to
$\hbress'\in\Bdset{k+1}{b}$.

\begin{lemma}\label{B1}
For $0\le b\le k$,
\begin{equation}\label{Eq:B1}
\BGF{k+1}{b}(a)=q^{a^2}\,\tBdbGF(a)
\end{equation}
\end{lemma}

\Proof
The proof of Lemma {\bf \ref{A1}} shows that the $\B$-transform provides
a bijection between $\tAdset{k}{b}(a)$ and $\Adset{k+1}{b}(a)$.
The same is thus true for $\tBdset{k}{b}(a)$ and $\Bdset{k+1}{b}(a)$.
The required result then immediately follows from Lemma {\bf \ref{A1param}}.
\cqfd

\begin{lemma}\label{B1GF}
If $0\le b<k$ then
\begin{subequations}\label{Eq:B1GF}
\begin{align}
\label{Eq:B1GFa}
\tBGF{k}{b}(a)&=\sum_{n=0}^a \frac{q^{n^2}}{(q;q)_{a-n}}\,\tBGF{k-1}{b}(n)
\\
\intertext{If $k>0$ then}
\label{Eq:B1GFb}
\tBGF{k}{k}(a)&=\sum_{n=0}^a \frac{q^{n(n+1)}}{(q;q)_{a-n}}\,\tBGF{k-1}{k-1}(n)
\end{align}
\end{subequations}
\end{lemma}

\Proof
The first expression here results from substituting the $k\to k-1$ case
of \eqref{Eq:B1} into \eqref{Eq:B2}.
The second expression results from first substituting
\eqref{Eq:U2} into the $b=k$ case of \eqref{Eq:B2},
and then applying the $k\to k-1$ and $b\to k-1$ case of \eqref{Eq:B1}.
\cqfd

\subsection{The sum side of the Bressoud identities}

Concatenating the above $\B$-transforms together yields 

\begin{theorem}\label{Bchain}
If $k\ge1$ and $0\le b\le k$ then
\begin{equation}\label{Eq:Bchain}
\tBGF{k}{b}(a)=
\sum_{a\ge n_k\ge n_{k-1}\ge \cdots\ge n_1\ge0}
\frac{q^{n_1^2+n_2^2+ \cdots+n_k^2
      +n_{1}+n_{2}+ \cdots+n_{b}}}
     {(q;q)_{a-n_k}(q;q)_{n_k-n_{k-1}}(q;q)_{n_{k-1}-n_{k-2}}
      \cdots(q;q)_{n_{2}-n_1}(q^2;q^2)_{n_1}}
\end{equation}
\end{theorem}

\Proof
Each $(\hbress,\ho)\in\tBdset{0}{0}(a)$ has $\hbress=(0,0,0, \cdots)$.
Because $\ho$ here is a partition with at most $a$ parts,
all of which are even,
$\tBGF{0}{0}(a)=(q^2;q^2)_a^{-1}$.
The two cases of \eqref{Eq:B1GF} then imply that
\begin{subequations}
\begin{align}
\label{Eq:BchainPf1a}
\tBGF{1}{0}(a)
&=\sum_{n_1=0}^a\frac{q^{n_1^2}}{(q;q)_{a-n_1}(q^2;q^2)_{n_1}}\,,\\
\label{Eq:BchainPf1b}
\tBGF{1}{1}(a)
&=\sum_{n_1=0}^a\frac{q^{n_1^2+n_1}}{(q;q)_{a-n_1}(q^2;q^2)_{n_1}}\,,
\end{align}
\end{subequations}

\noindent which are the $k=1$ cases of \eqref{Eq:Bchain}.
The proof of the general case then proceeds exactly as the proof of
Theorem {\bf \ref{Achain}}.
\cqfd

\noindent
Note that the RHS of \eqref{Eq:Bchain} is identical with the
$i=k-b$ case of \eqref{Eq:CPAGc},
as is seen on setting $n_j\to\lambda_{k+1-j}$.

\subsection{Characterising decorated even $\B$-paths}
\label{CharPathE}

Because $\tBdb(a)$ is a subset of $\tAdb(a)$,
the conditions \eqref{Eq:DeltaCons} on
the points $\{(i_j,b_j)\}_{j=1}^a$ of $\thbress\in\tBdb(a)$
continue to apply.
The extra condition that applies to $\tBdb(a)$
is that, for $1\le j\le a$,
\begin{equation}\label{Eq:DeltaConsE}
b_j=k\implies i_j\equiv k-b\pmod2
\end{equation}

\noindent Each set of points $\{(i_j,b_j)\}_{j=1}^a$ that satisfies
\eqref{Eq:DeltaCons} and \eqref{Eq:DeltaConsE} determines a unique 
element $\thbress\in\tBdb(a)$.

\subsection{Bijection from cylindric partitions to decorated even $\B$-paths}
\label{BijDEBpaths}

For $d$ even, set $d=2k$. Let $\Lambda=[d-x,x]\in\wtlatd{2}{d}$.
For now, we only consider $x\le k$. Set $b=k-x$, whereupon $0\le b\le k$.
In what follows, for $\pi\in\CPX{2}{\Lambda}(a)$, we use
$\boldLambda(\pi)=(\cdots,\Lambda^{(3)},\Lambda^{(2)},\Lambda^{(1)})$
and $\bolddelta(\pi)=(\cdots,\delta_3,\delta_2,\delta_1)$
to construct an element $\thbress\in\tBdset{k}{b}(a)$
by specifying the set of points $\{(i_j,b_j)\}_{j=1}^a$
of its $a$ peaks and deaks.
We then show that this defines a weight-preserving bijection between
$\CPX{2}{\Lambda}(a)$ and $\tBdset{k}{b}(a)$.

As in Section {\bf \ref{BijDBpaths}},
set $i_j=\sum_{m=j}^a \delta_m$ for $1\le j\le a$, set $i_{a+1}=0$,
and, for $j>0$, let $x_j$ be such that $\Lambda^{(j)}=[d-x_j,x_j]$,
and obtain $b_j$ using \eqref{Eq:bjDef}.
In particular, $b_j=b$ for $j>a$.
Because the procedure of Section {\bf \ref{BijDBpaths}} has already
been shown to lead to an element $\thbress\in\tAdset{k}{b}(a)$, it is
only necessary to check, additionally, that \eqref{Eq:DeltaConsE} holds.
First, we claim that

\begin{equation}\label{Eq:LogicClaim}
x_j\le k\quad\Leftrightarrow\quad i_j\equiv b_j-b\pmod2,
\end{equation}

\noindent for $1\le j\le a+1$.
That this holds for $j=a+1$ is immediate.
Because $i_{j+1}=i_{j}-\delta_j$, the $j\to j+1$ case of
\eqref{Eq:LogicClaim} implies that
$x_{j+1}\le k\Leftrightarrow i_{j}\equiv \delta_j+b_{j+1}-b\pmod2$.
Thereupon, \eqref{Eq:LogicClaim} itself is obtained from \eqref{Eq:Logic},
and the claim is proved by induction.
Now, if $b_j=k$ then \eqref{Eq:bjDef} implies that $x_j=0$,
whereupon the above claim gives $i_j\equiv k-b\pmod2$, as required.

To show that the map from $\CPX{2}{\Lambda}$ to $\tBdset{k}{b}(a)$
so described is a bijection, first note that
given values $\{(i_j,b_j)\}_{j=1}^a$,
the argument of Section {\bf \ref{BijDBpaths}} uniquely determines
values $x_a, x_{a-1},$ $\cdots$, $x_1$, for which $0 \le x_j \le 2k+1$ 
for $1\le j\le k$. However, here we require each $x_j\le 2k$.
We must show that $x_j=2k+1$ is not possible.
Indeed, $x_j=2k+1$ can only arise from \eqref{Eq:bjDef} if $b_j=k$.
However, for such a case,
\eqref{Eq:DeltaConsE} combined with \eqref{Eq:LogicClaim} would give
$x_j\le k$, and then from \eqref{Eq:bjDef} that $x_j=0$.
This establishes the bijection, and, as in Section {\bf \ref{BijDBpaths}},
it is weight-preserving.

\subsection{The product side of the Bressoud identities}

From the above bijection, we obtain

\begin{lemma}\label{WtPresEven}
For $k\ge0$, let $\Lambda\in\wtlatd{2}{2k}$
be of the form $\Lambda=[k+b,k-b]$ for $0\le b\le k$.
Then, for each $a\ge0$,

\begin{equation}\label{Eq:WtPresEven}
\CPgfX{2}{\Lambda}(a)=\tBGF{k}{b}(a)
\end{equation}
\end{lemma}

\noindent Using this result, the even $d$ case of Theorem {\bf \ref{CPAG}}
then follows from Theorem {\bf \ref{Bchain}},
after setting $b=k-i$ and $n_j=\lambda_{k+1-j}$.

Note that the above analysis omits the cases for which
$\Lambda=[k-b,k+b]$ for $0<b\le k$.
From the point of view of the cylindric partitions,
it is easy to see that
$\CPgfX{2}{[k-b,k+b]}(a)=\CPgfX{2}{[k+b,k-b]}(a)$,
and thus \eqref{Eq:WtPresEven} leads to sum-type expressions for these cases.
On the other hand, the above bijection into decorated even $\B$-paths
extends naturally so that this case maps into an analogous set of
decorated paths $\thbress=(\hbress,\ho)$
for which the constraint on $\hbress=(h_0,h_1,h_2, \cdots)$
that if $h_i=k$ then $i\equiv k-b\pmod2$,
is replaced by the constraint that if $h_i=k$ then $i\not\equiv k-b\pmod2$.
However, applying the methods of the previous sections
to these paths (starting with the $k=1$ case instead of $k=0$),
leads to sum-type expressions for their generating functions
that are exactly those obtained from \eqref{Eq:WtPresEven} and
the equality $\CPgfX{2}{[k-b,k+b]}(a)=\CPgfX{2}{[k+b,k-b]}(a)$.

\subsection{The interpretation of the proof in terms of abaci}
In the even-level case, an interpretation for the expression
\eqref{Eq:Bchain} for the generating function $\tBGF{k}{b}(a)$
in terms of moves of the yokes of the corresponding abaci can 
be given that is very similar to that described in section {\bf \ref{YokeMoves}} 
for the odd-level cases. One difference to the description given there is that 
if there is a peak or deak at $j$ then the weight of the corresponding yoke depends 
on $\hh_j$: it is either $[i+\hh_j,i-\hh_j]$ or $[i-1-\hh_j,i+1+\hh_j]$,
with the choice between these two determined by \eqref{Eq:rDist}
(the weight of each yoke in the Dirac sea is $[i+b,i-b]$).

\section{Discussion}
\label{discussion}

\subsection{Summary of results}
In this paper, we have brought together known results on product expressions
for the characters of $\slchap{r}$, $\cW_r$, and the generating functions of 
cylindric partitions, giving simple proofs of each, and showing how they are 
related. In addition, we have applied combinatorial methods to the cylindric 
partitions in the $r=2$ case to obtain non-negative sum expressions 
for their generating functions. Equating the sum and the product expressions,
and cancelling a common factor, we retrieved the known Andrews-Gordon-Bressoud
extensions of the Rogers-Ramanujan identities.

\subsection{Higher-rank identities}
The methods of the current papers can be extended to investigate 
Rogers--Ramanujan-type identities for other affine Lie algebras.
In particular, as we will show elsewhere,
simple proofs of the $\cW_3$ Rogers--Ramanujan-type identities of
\cite{andrews.schilling.warnaar.1999} 
\footnote{\,
See also \cite{feigin.foda.welsh.2008}.
}
can be obtained.

\subsection{An intermediate set of ordinary partitions} 
In the proof of the sum side of the Andrews-Gordon identities,
a term $(q;q)_{a-\lambda_1}^{-1}$ appears in the summand
(see \eqref{Eq:CPAG}),
in addition to and on equal footing with the terms that must appear in 
any proof of these identities.
A similar additional term appears in the proof of the product side 
of the Andrews-Gordon identities.
In the limit $a \rightarrow \infty$, these extra terms become 
$(q;q)^{-1}_{\infty}$, the generating function of ordinary partitions,
and cancel in the final result, leading to the known identities. 

Though they eventually cancel, the presence of these terms is crucial
in a proof based on cylindric partitions.
In a sense, this is the main lesson learnt of Burge's work
on Rogers-Ramanujan-type identities \cite{burge.1993}.
Extending the set of combinatorial objects that we work with,
by introducing a set of ordinary partitions into the mix,
greatly simplifies the proofs.
The same factor appeared in
\cite{andrews.schilling.warnaar.1999}.
A similar remark applies to the proof of the Bressoud identities.

\subsection{Cylindric partitions in other contexts}
The cylindric partitions that appear in this work have also appeared in Postnikov's work on quantum 
Schubert calculus \cite{postnikov.2002}
\footnote{\,
We thank Ch Krattenthaler for bringing Postnikov's work to our attention.
}. 
More recently, cylindric partitions have appeared in AGT-type computations of conformal blocks in minimal 
conformal field theories based on $\cW_r$ times a $U(1)$ factor 
\cite{belavin.foda.santachiara}, \cite{alkalaev.belavin}--\cite{fukuda.nakamura.matsuo.zhu}. This relation 
was discussed briefly in subsection {\bf \ref{r.burge.partitions}}. 
In this context, the $(q;q)_{\infty}^{-1}$ factor that appears in our derivations of the Andrews-Gordon-Bressoud
identities, in section {\bf \ref{andrews.gordon}} and section {\bf \ref{bressoud}}, and that cancels so that we 
end up with the known forms of these identities, is the result of the action of a Heisenberg algebra whose presence 
is crucial to the derivation of amenable expressions for the conformal blocks. 

\begin{appendix}

\section{The $\slchap{r}$ Macdonald identity}
\label{macdonald.identity}

The $\slchap{r}$ Macdonald identity \cite{macdonald.1972}
may be expressed in the form (\cite[Theorem 1.61]{milne.1985})
\begin{equation}\label{Eq:MacId1}
\sum_{\sigma\in\mathfrak S_r}
(-1)^{\ell(\sigma)}
\sum_{k_1+ \cdots+k_r=0}\,
\prod_{i=1}^r x_{\sigma(i)}^{rk_{\sigma(i)}+i-{\sigma(i)}}
              q^{\frac12rk_{\sigma(i)}^2+ik_{\sigma(i)}}
=(q;q)^{r-1}_{\infty}
\prod_{1\le i<j\le r} \ll \frac{x_i}{x_j},q\frac{x_j}{x_i}; q \rr_{\infty},
\end{equation}

\noindent where $\mathfrak S_r$ is the symmetric group
and $\ell(\sigma)$ is the length of $\sigma\in\mathfrak S_r$.
Expressing the signed sum over $\mathfrak S_r$ as a determinant,
leads to the following form of the identity:
\begin{equation}\label{Eq:MacId3}
\sum_{k_1+ \cdots+k_r=0}
\det_{1\le s,t\le r}
\ll x_t^{rk_t+s-t}q^{\frac12rk_t^2+sk_t} \rr \\
=(q;q)^{r-1}_{\infty}
\prod_{1\le i<j\le r} \ll \frac{x_i}{x_j},q\frac{x_j}{x_i}; q \rr_{\infty}
\end{equation}

\noindent We can also express this identity in a slightly different form.
Renaming each $k_{\sigma(i)}$ by $k_i$ in the second sum of \eqref{Eq:MacId1},
then exchanging the order of the two sums,
and finally expressing the signed sum over $\mathfrak S_r$ as a determinant,
leads to the following:
\begin{equation}\label{Eq:MacId2}
\sum_{k_1+ \cdots+k_r=0}
\det_{1\le s,t\le r}
\ll x_t^{rk_s+s-t}q^{\frac12rk_s^2+sk_s} \rr \\
=(q;q)^{r-1}_{\infty}
\prod_{1\le i<j\le r} \ll \frac{x_i}{x_j},q\frac{x_j}{x_i}; q \rr_{\infty}
\end{equation}

\section{$\slchap{r}$ weight space and characters}
\label{affine.characters}

\subsection{Cartan data}
\label{Cartan}

Let $\{\alpha_i\}_{i=0}^{r-1}$ be the simple roots of $\slchap{r}$
and $\{\Lambda_i\}_{i=0}^{r-1}$ the corresponding fundamental weights.
The dual $\cartand$ of the Cartan subalgebra $\cartan$ of $\slchap{r}$
has basis $\{\Lambda_0,\alpha_0,\alpha_1, \cdots,\alpha_{r-1}\}$.
An alternative basis is
$\{\delta,\Lambda_0,\Lambda_1, \cdots,\Lambda_{r-1}\}$,
where the \emph{null root} $\delta$ is defined by
$\delta=\sum_{i=0}^{r-1}\alpha_i$.
The \emph{Weyl vector} $\rho$ is defined by $\rho=\sum_{i=0}^{r-1}\Lambda_i$.
The above bases are related by introducing orthonormal vectors
$\epsilon_1, \epsilon_2, \cdots, \epsilon_r$
and setting
\footnote{\, 
Although much of this material is similar to that in
Section 2 of \cite{jimbo.miwa.okado.1988},
our $\epsilon_i$ corresponds to their $\epsilon_{i-1}$.
Note that in view of \eqref{Eq:EpsilonExps1}, our system is
such that the difference between the coefficients of
$\epsilon_i$ and $\epsilon_r$ is equal to the length
of the $i$th row of the corresponding partition $\parop(\Lambda)$,
defined below.
}

\begin{subequations}\label{Eq:CartanDefs}
\begin{align}
\label{Eq:CartanDefsa}
\Lambda_i&=\epsilon_1+ \cdots+\epsilon_{i}-i\epsilon+\Lambda_0
\qquad(0<i<r),\\
\label{Eq:CartanDefsb}
\alpha_0&=\epsilon_{r}-\epsilon_{1}+\delta,
\quad
\alpha_i=\epsilon_{i}-\epsilon_{i+1}
\quad(0<i<r),
\end{align}
\end{subequations}

\noindent with $\epsilon=\frac1r\sum_{j=1}^{r}\epsilon_j$.
To each $\Lambda=\sum_{i=0}^{r-1}m_i\Lambda_i\in\wtlatd{r}{d}$,
there is a corresponding partition
\footnote{\, 
This partition is the tableau signature $(f_1, \cdots,f_r)$
described in Section 2 of \cite{jimbo.miwa.okado.1988}, when $f_r=0$.
There, adding or removing a column of length $r$ doesn't change
the corresponding element of $\wtlatd{r}{d}$.
}
$\parop(\Lambda)=(\mu_1, \cdots,\mu_{r-1})\in\partset{r-1}$
defined by $\mu_j=\sum_{i=j}^{r-1}m_i$ for $j=1, \cdots,r-1$.
Note that $\mu_1\le d$ (and $\mu_r=0$).
Then \eqref{Eq:CartanDefsa} gives

\begin{subequations}
\begin{align}
\label{Eq:EpsilonExps1}
\Lambda&=d\Lambda_0
         +\sum_{i=1}^{r} \mu_i\epsilon_i
         -\epsilon\sum_{i=1}^{r}\mu_i,\\
\label{Eq:EpsilonExps2}
\rho&=r\Lambda_0+\sum_{i=1}^r (r-i)\epsilon_i-\epsilon\sum_{i=1}^r (r-i)
=r\Lambda_0-\sum_{i=1}^r i\epsilon_i+\epsilon\sum_{i=1}^r i,\\
\label{Eq:EpsilonExps3}
\Lambda+\rho&=(r+d)\Lambda_0
              +\sum_{i=1}^{r} (\mu_i-i)\epsilon_i
              -\epsilon\sum_{i=1}^{r}(\mu_i-i)
\end{align}
\end{subequations}

An inner product $\wtinner{\cdot}{\cdot}$ on $\cartand$ is defined by,
in addition to $\wtinner{\epsilon_i}{\epsilon_j}=\delta_{ij}$, setting

\begin{equation}
\wtinner{\delta}{\Lambda_0}=1,\qquad
\wtinner{\delta}{\delta}=\wtinner{\Lambda_0}{\Lambda_0}=
\wtinner{\delta}{\epsilon_i}=\wtinner{\Lambda_0}{\epsilon_i}=0
\end{equation}

\noindent This implies that

\begin{equation}
\wtinner{\Lambda_i}{\Lambda_j}=\min\{i,j\}-\frac{ij}{r},\qquad
\wtinner{\alpha_i}{\Lambda_j}=\delta_{ij},\qquad
\wtinner{\alpha_i}{\alpha_j}=2\delta_{ij}
                     -\delta_{i+1,j}-\delta_{i,j+1},\qquad
\end{equation}

\noindent for $0\le i,j<r$, where $\delta_{ij}=1$ if $i=j$, and 
$\delta_{ij}=0$ otherwise. Additionally, we obtain

\begin{equation}
\wtinner{\epsilon}{\epsilon_j}=\wtinner{\epsilon}{\epsilon}=\frac1r,
\qquad
\wtinner{\alpha_i}{\epsilon}=0
\end{equation}

For any affine Lie algebra $\lieg$ the Weyl group $W$ of $\lieg$ is the 
group generated by \emph{Weyl reflections} $\{s_\alpha\}_{\alpha\in\Pi}$, 
where $\Pi$ is the set of simple roots, which act on $\Lambda\in\cartand$ 
according to

\begin{equation}\label{Eq:WeylReflect}
s_{\alpha}(\Lambda)=
\Lambda-2\frac{\wtinner{\Lambda}{\alpha}}{\wtinner{\alpha}{\alpha}}
\alpha
\end{equation}

\noindent For the simple root $\alpha_i\in\Pi$, set $s_i=s_{\alpha_i}$.
If $w=s_{i_1}s_{i_2} \cdots s_{i_\ell}$, and $w$ cannot be written
as a shorter product of the generators,
then we say that $\ell$ is the \emph{length} of $w$ and write $\ell(w)=\ell$.

For any affine Lie algebra $\lieg$, $W$ can be written as a semi-direct product

\begin{equation}\label{Eq:SemiDirect}
W=T \ltimes \cW_0,
\end{equation}

\noindent where $T$ is an infinite abelian group and $\cW_0$ is the
Weyl group of the corresponding simple Lie algebra $\lieg_0$
\cite[Section 3]{kac.book.1990}.
With $\cartand_0\subset\cartand$ the weight space of $\lieg_0$,
the group $T$ is generated by certain elements $t_{\alpha}$,
indexed by $\alpha\in\cartand_0$, that act on $\Lambda\in\cartand$
according to

\begin{equation}\label{Eq:TranAction}
t_{\alpha}(\Lambda)=\Lambda+
\wtinner{\Lambda}{\delta}\alpha
-
\ll
\wtinner{\Lambda}{\alpha} + \tfrac12 \wtinner{\alpha}{\alpha} \wtinner{\Lambda}{\delta}
\rr 
\delta
\end{equation}

\noindent It can be shown that each $\ell(t_{\alpha})\in2\ZZp$.

In the $\lieg=\slchap{r}$ case,
$T=\{t_{\alpha}\}_{\alpha\in\rootlatc{r}}$ where $\rootlatc{r}$
is the root lattice of $\slclass{r}$, given explicitly by
$\rootlatc{r}=
\{n_1\alpha_1+n_2\alpha_2+ \cdots+n_{r-1}\alpha_{r-1}\,|\,n_i\in\ZZ\}$.
Alternatively, because each $\alpha_i=\epsilon_i-\epsilon_{i+1}$,

\begin{equation}\label{Eq:RootLat}
\rootlatc{r}=
\{k_1\epsilon_1+k_2\epsilon_2+ \cdots+k_{r}\epsilon_{r}\,|\,
k_i\in\ZZ,\,k_1+k_2+ \cdots+k_{r}=0\}
\end{equation}

\noindent Also in the case $\lieg=\slchap{r}$, $\cW_0\cong\mathfrak S_r$
and this acts naturally on the indices of the $\epsilon_i$ in that
$\sigma(\epsilon_i)=\epsilon_{\sigma(i)}$ for all $\sigma\in\mathfrak S_r$.
After defining $\sigma(\delta)=\delta$ and $\sigma(\Lambda_0)=\Lambda_0$,
the action of $\mathfrak S_r$ is extended linearly to the whole of $\cartand$.

\subsection{Affine characters}
\label{AffChars}

For an affine Lie algebra $\lieg$, the character $\chi^{\lieg}_\Lambda$ of 
an integrable highest weight $\lieg$-module of highest weight $\Lambda$ is 
given by the following theorem

\begin{theorem}\label{FullChar}
\cite[Section 10.4]{kac.book.1990}
\begin{equation}\label{Eq:CharDef}
\chi^{\lieg}_\Lambda
=
e^\Lambda\frac{{\mathcal N}^{\lieg}_\Lambda}{{\mathcal N}^{\lieg}_0},
\end{equation}

\noindent where

\begin{equation}\label{Eq:NumDef}
{\mathcal N}^{\lieg}_\Lambda=
\sum_{w\in W} (-1)^{\ell(w)} e^{w(\Lambda+\rho)-(\Lambda+\rho)},
\end{equation}

\noindent in which $W$ is the Weyl group of $\lieg$.
\end{theorem}

In the $\slchap{r}$ case, noting \eqref{Eq:SemiDirect},
each element $w\in W$ is of the form $w=t_{\alpha}\sigma$
for $\alpha=\sum_{i=1}^r{k_i\epsilon_i}$
with $\sum_{i=1}^r{k_i}=0$, and $\sigma\in\mathfrak S_r$.
For such $w$, use of \eqref{Eq:EpsilonExps3} and \eqref{Eq:TranAction}
yields

\begin{multline}
\label{Eq:WeylAction}
w(\Lambda+\rho)-(\Lambda+\rho)
= \ll t_{\alpha}(\sigma(\Lambda+\rho))-\sigma(\Lambda+\rho) \rr + \sigma(\Lambda+\rho)-(\Lambda-\rho)
\\
=(r+d)\sum k_i\epsilon_i - \ll \sum(\mu_i-i)k_{\sigma(i)} +\tfrac12(r+d)\sum k_i^2 \rr \delta
\\
+\sum \ll \mu_i-i-\mu_{\sigma(i)}+\sigma(i) \rr \epsilon_{\sigma(i)},
\end{multline}

\noindent where each summation is over $1\le i\le r$.
On setting $x_i=e^{-\epsilon_i}$ and $q=e^{-\delta}$,
we then obtain

\begin{equation}\label{Eq:ExpAction}
e^{w(\Lambda+\rho)-(\Lambda+\rho)}
=
\prod_{i=1}^r
x_{\sigma(i)}^{-(r+d)k_{\sigma(i)}-\mu_i+i+\mu_{\sigma(i)}-\sigma(i)}
q^{ (\mu_i-i)k_{\sigma(i)} +\frac12(r+d)k_{\sigma(i)}^2 }
\end{equation}

\noindent Note that $(-1)^{\ell(w)}=(-1)^{\ell(\sigma)}$.
Substituting \eqref{Eq:ExpAction} into \eqref{Eq:NumDef} and writing
the sum as a sum over $k_1, \cdots,k_r$ with $k_1+ \cdots+k_r=0$, and 
a signed sum over $\sigma\in\mathfrak S_r$, and then expressing the 
latter sum as a determinant, we obtain

\begin{equation}\label{Eq:Num}
{\mathcal N}^{\slchap{r}}_\Lambda=
\sum_{k_1+ \cdots+k_r=0}
\det_{1\le s,t\le r}
\ll x_{s}^{-(r+d)k_{s}-\mu_t+t+\mu_{s}-s}
q^{ (\mu_t-t)k_{s} +\frac12(r+d)k_s^2 } \rr
\end{equation}

\noindent In the $\Lambda=0$ case, we find that

\begin{equation}\label{Eq:DenId}
{\mathcal N}^{\slchap{r}}_0
=
\sum_{k_1+ \cdots+k_r=0}
\det_{1\le s,t\le r}
\ll x_{s}^{-rk_{s}+t-s}
q^{ -tk_{s} +\frac12rk_s^2 } \rr
=
(q;q)^{r-1}_{\infty}
\prod_{1\le i<j\le r} \ll \frac{x_i}{x_j},q\frac{x_j}{x_i}; q \rr_{\infty},
\end{equation}

\noindent the final equality resulting from the
Macdonald identity \eqref{Eq:MacId3} after noting that the sign
of each $k_s$ in the sum can be changed.
This identity \eqref{Eq:DenId} is the \emph{denominator identity}
for $\slchap{r}$ \cite[eqn.~(10.4.4)]{kac.book.1990}.

In Section {\bf \ref{SlrChars}} of the main text, we evaluate the principal specialisations 
of $\chi^{\lieg}_\Lambda$ and ${\mathcal N}^{\slchap{r}}_\Lambda$. These are the specialisations 
for which $e^{\Lambda}\to1$ and $e^{-\alpha_i}\to q$ for each $\alpha_i\in\Pi$. 
Because $\delta=\sum_{i=0}^{r-1}\alpha$, we also have $e^{-\delta}\to q^r$. Note that, because 
$\alpha_i=\epsilon_i-\epsilon_{i+1}$ and $x_i=e^{-\epsilon_i}$, the specialisation 
$e^{-\alpha_i} \to q$ is effected by $x_i\to q^{-i}$.
\footnote{\,
Note that only ratios $x_i/x_j$ are required.
}

\subsection{$\cW_r$ characters}
\label{WrNotes}

For $r\le p<p'$ with $p$ and $p'$ coprime, the $\cM^{\, p, \pp}_r$ minimal model 
character of the highest weight $\cW_r$ representation labelled by the pair
$\xi \in \wtlatd{r}{p-r}$ and $\zeta \in \wtlatd{r}{p'-r}$ is given by 
\cite{mizoguchi.1991}--\cite{nakanishi.1990},

\begin{equation}\label{Eq:WrCharApp}
\chi^{r,p,p'}_{\xi,\zeta}=
\frac{1}{\eta(q)^{r-1}}
\sum_{\alpha\in\rootlatc{r}}\sum_{\sigma\in\mathfrak S_r}
(-1)^{\ell(\sigma)}\,
q^{\frac12pp'|\alpha-(\xi+\rho)/p+\sigma(\zeta+\rho)/p'|^2}
\end{equation}

\noindent Here, $\eta(q)=q^{1/24}(q;q)_\infty$ is the Dedekind $\eta$-function,
and $\rootlatc{r}$ is given by \eqref{Eq:RootLat}.

We now express \eqref{Eq:WrCharApp} in terms of $\mu$ and $\nu$,
defined by
$\mu=(\mu_1,\mu_2, \cdots,\mu_r)=\parop(\zeta)$ and
$\nu=(\nu_1,\nu_2, \cdots,\nu_r)=\parop(\xi)$.
For the exponent of \eqref{Eq:WrCharApp},

\begin{multline}
\label{Eq:WrCharExppp}
\Biggl| \alpha-\frac{\xi+\rho}{p}
                   +\frac{\sigma(\zeta+\rho)}{p'}\Biggr|^2
=
|\alpha|^2
-\frac{2}{p}\wtinner{\alpha}{\xi+\rho}
+\frac{2}{p'}\wtinner{\sigma^{-1}(\alpha)}{\zeta+\rho}
\\
-\frac{2}{pp'}\wtinner{\sigma^{-1}(\xi+\rho)}{\zeta+\rho}
+\Biggl|\frac{\xi+\rho}{p}\Biggr|^2
+\Biggl|\frac{\zeta+\rho}{p'}\Biggr|^2
\end{multline}

\noindent From \eqref{Eq:RootLat}, each $\alpha\in\rootlatc{r}$ is of
the form $\alpha=\sum_{i=1}^r k_i\epsilon_i$ with $\sum_{i=1}^r k_i=0$.
Then $\sigma^{-1}(\alpha)=\sum_{i=1}^r k_{\sigma(i)}\epsilon_i$.
After writing

\begin{equation}
\wtinner{\sigma^{-1}(\xi+\rho)}{\zeta+\rho}
=\wtinner{\sigma^{-1}(\xi+\rho)-(\xi+\rho)}{\zeta+\rho}
+\wtinner{\xi+\rho}{\zeta+\rho},
\end{equation}

\noindent and using the expression \eqref{Eq:EpsilonExps3}
for $\xi+\rho$ and $\zeta+\rho$, we obtain

\begin{multline}
\label{Eq:WrCharExp}
\frac12pp'
\Biggl|\alpha-\frac{\xi+\rho}{p} +\frac{\sigma(\zeta+\rho)}{p'}\Biggr|^2
=
\\
p'\sum_{i=1}^r k_i(\tfrac12pk_i-\nu_i+i)
+\sum_{i=1}^r (pk_{\sigma(i)}-\nu_{\sigma(i)}+\sigma(i)+\nu_i-i)(\mu_i-i)
+\hat\Delta^r_{\xi,\zeta},
\end{multline}

\noindent where we define

\begin{equation}\label{Eq:DeltaDef}
\hat\Delta^r_{\xi,\zeta}=
\frac12pp'
\Biggl|\frac{\xi+\rho}{p}-\frac{\zeta+\rho}{p'}\Biggr|^2
\end{equation}

\noindent Substituting this into \eqref{Eq:WrCharApp},
writing the signed sum over $\mathfrak S_r$ as a determinant,
and using $\eta(q)=q^{1/24}(q;q)_\infty$,
yields

\begin{equation}\label{Eq:WrChar2}
\begin{split}
\chi^{r,p,p'}_{\xi,\zeta}
&=
\frac{q^{\hat\Delta^r_{\xi,\zeta}-(r-1)/24}}{(q;q)^{r-1}_\infty}
\sum_{k_1+ \cdots+k_r=0}
q^{p'\sum_{i=1}^r k_i(\frac12 pk_i-\nu_i+i)}
\det_{1\le s,t\le r}
\ll q^{(\mu_t-t)(pk_s-\nu_s+s+\nu_t-t)} \rr
\end{split}
\end{equation}

\section{Kyoto patterns and cylindric partitions}
\label{Kyoto}

In \cite[Section 3]{jimbo.misra.miwa.okado.1991}, the authors define a \emph{pattern}
to be a two-index array of integers $t_{jk}$ for $j\in\ZZ$ and $k\in\ZZp$,
which is subject to:
\begin{enumerate}
\item For each $j$, there exists $\gamma_j$ such that
      the sequence $\{\gamma_j-t_{jk}\}_{k\ge0}$ is a partition;
\item $t_{jk}\le t_{j+1,k}$ for all $j$ and $k$;
\item $t_{j+d,k}=t_{jk}+r$ for all $j$ and $k$.
\end{enumerate}
Such a pattern is said to be \emph{normalised} if
$0\le\gamma_1\le\gamma_2\le \cdots\le\gamma_d<r$.

We will now show that such a normalised sequence encodes
a cylindric partition of type $(\infty^r)/\mu/d$
where the partition $\mu=(\mu_1,\mu_2, \cdots,\mu_{r})$
is defined so that its conjugate $\mu'$ is given by
$\mu'=(\gamma_d-\gamma_1,\gamma_d-\gamma_2, \cdots,\gamma_{d}-\gamma_{d-1},0)$.
The $j$th column of the corresponding plane partition $\pi$ is
then obtained from the sequence $\{t_{jk}\}_{k\ge0}$
by, for $k>0$, placing $t_{jk}-t_{j,k-1}$ entries $k$ in that column.
In other words, the values in the $j$th column are the parts
of the partition conjugate to $\{\gamma_j-t_{jk}\}_{k\ge0}$.
It is easily checked that the cylindric partition $\pi$
so defined satisfies the conditions \eqref{Eq:PlaneCons}.

For $1\le j\le d$, let $\lambda^{(j)}$ denote the partition
$\{\gamma_j-t_{jk}\}_{k\ge0}$.
The array $\{t_{jk}\}$ is then also encoded in the multipartition
$\boldlambda=(\lambda^{(1)},\lambda^{(2)}, \cdots,\lambda^{(d)})$.
Because these multipartitions inherit a cylindrical embedding property,
they are termed cylindrical multipartitions
(in \cite{jimbo.misra.miwa.okado.1991,foda.leclerc.okado.thibon.welsh} the boxes of $\boldlambda$ are coloured
in a certain periodic manner).
Such multipartitions were first considered in \cite{date.jimbo.kuniba.miwa.okado.1989}.

We thus see that translating between a normalised pattern $\{t_{jk}\}$,
a cylindric partition and a cylindrical multipartition is
straightforward.
For example, consider the cylindrical multipartition of
Example 2.12 of \cite{foda.leclerc.okado.thibon.welsh}
for which $r=4$, $d=3$ and $\Lambda=[0,0,1,2]$.
\footnote{\, 
In \cite{foda.leclerc.okado.thibon.welsh}, this is labelled by $[2,1,0,0]\in\wtlatd43$. However, 
to agree with the conventions of the current paper, it should be labelled 
by $[0,0,1,2]$,
or any cyclic permutation of the indices therein.
}
Here $\boldlambda=\bigl((10,10,8,4,4),(9,9,1,1),(10,7,1)\bigr)$.
The description above then leads to the plane partition

\begin{equation}\label{Eq:TypCPP3}
{\scriptsize
\begin{tabular}{cccccccccccccccccc}
&&&&&&&&10&4&1&$\cdot$&$\cdot$\\
&&&&&&10&9&7&4&$\cdot$&$\cdot$\\
&&&&&&10&9&1&$\cdot$&$\cdot$\\
&&&&&&8&1&$\cdot$&$\cdot$\\
&&&&&10&4&1&$\cdot$&$\cdot$\\
&&&10&9&7&4&$\cdot$&$\cdot$\\
&&&10&9&1&$\cdot$&$\cdot$\\
&&&8&1&$\cdot$&$\cdot$\\
&&10&4&1&$\cdot$&$\cdot$\\
10&9&7&4&$\cdot$&$\cdot$\\
10&9&1&$\cdot$&$\cdot$\\
8&1&$\cdot$&$\cdot$\\
\end{tabular}
}%
\end{equation}

The patterns $\{t_{jk}\}$ that pertain to irreducible characters
of $\slchap{r}$ require a further constraint.
Proposition 3.4 of \cite{jimbo.misra.miwa.okado.1991} designates a normalised pattern
$\{t_{jk}\}$ as \emph{highest-lift} if,
for each $k>0$, there exists $j$ such that $t_{j+1,k-1}>t_{jk}$.
Through the correspondence between patterns and plane
partitions described above, this condition translates to
state that for each $k>0$, there exists at least one row of $\pi$
in which no entry $k$ appears.
This is also equivalent to the condition on multipartitions
described in \cite[Proposition 2.11]{foda.leclerc.okado.thibon.welsh}.
\footnote{\,
If $\mu=\hat\mu(\Lambda)$,
the colouring of the boxes enables the full
character $\chi^{\slchap{r}}_\Lambda$ to be obtained by
enumerating the highest-lift multipartitions.
}
Multipartitions that respect this condition index the irreducible representations 
of the Ariki-Koike algebras $\mathcal H(v;u_1, \cdots,u_d)$, in the cases where each 
parameter is a certain root of unity \cite{ariki.1996}. The connection between this 
topic and the crystal basis theory of $U_q(\slchap{r})$ was elucidated and exploited 
in \cite{foda.leclerc.okado.thibon.welsh}.

\end{appendix}

\section*{Acknowledgements}
We thank Christian Krattenthaler for discussions of W H Burge's work, that took place almost 
twenty years ago. 
We obtained the bijection that we used in this work to derive the sum-side of the 
Andrews-Gordon-Bressoud identities while attempting to reconstruct Christian's explanations. 
OF wishes to thank S Corteel for discussions on the topic of this work, part of which was 
carried out during a visit to the 
{\it Laboratoire d'Informatique Algorithmique: Fondements et Applications [LIAFA], 
Universit\'e Paris Diderot 7}. 
He wishes to thank S Corteel and {\it LIAFA} for kind hospitality and the
{\it Fondation Sciences Mathematiques de Paris} for financial support during his visit. 
Both authors were supported by the Australian Research Council. 


\end{document}